\documentclass[letterpaper,10pt,twocolumn]{IEEEtran}  % Comment this line out
\IEEEoverridecommandlockouts                              % This command is only
\usepackage{mathptmx} % assumes new font selection scheme installed
\usepackage{amsmath} % assumes amsmath package installed
\usepackage{amssymb}  % assumes amsmath package installed
\usepackage{slashbox}
\usepackage{graphicx}
\usepackage{dsfont}
\usepackage{mathtools}
\usepackage{color}
\usepackage{ntheorem}
\usepackage{mathtools}
\usepackage{epstopdf}
\newtheorem{thm}{Theorem}[section]

\newtheorem{lem}[thm]{Lemma}
\newtheorem{prop}[thm]{Proposition}

\newtheorem{conj}[thm]{Conjecture}
\newtheorem{fact}[thm]{Fact}
\newtheorem{defn}[thm]{Definition}
\newtheorem{rem}[thm]{Remark}
\newcommand{\eps}{\epsilon}

\newcommand{\md} {\mathds}
\newcommand{\ra} {\rightarrow}

\title{Phase limitations of Zames-Falb multipliers}
\author{Shuai Wang, Joaquin Carrasco, and William  P. Heath 
         \thanks{School of Electrical and Electronic Engineering,
         The University of Manchester,
          Manchester M13 9PL, UK}
      \thanks{DOI: 10.1109/TAC.2017.2729162} 
      \thanks{\copyright 2017 IEEE. Personal use of this material is permitted. Permission from IEEE must be obtained for all other uses, in any current or future media, including reprinting/republishing this material for advertising or promotional purposes, creating new collective works, for resale or redistribution to servers or lists, or reuse of any copyrighted component of this work in other works.}
         \thanks{{\tt\small shuai.wang@manchester.ac.uk}}
         \thanks{{\tt\small joaquin.carrascogomez@manchester.ac.uk}}
         \thanks{{\tt\small william.heath@manchester.ac.uk}}
}

\begin{document}
\maketitle
\thispagestyle{empty}
\pagestyle{empty}
\begin{abstract}

%We derive phase limitations of discrete-time Zames-Falb multipliers. In some cases the phase limitations may be used to show there exists no appropriate Zames-Falb multiplier for a Lur'e system with a given LTI plant and a given class of slope-restricted nonlinearities. We conjecture that this in turn implies the Lur'e system is not absolutely stable.  We illustrate the results with some numerical examples and discuss the implications for benchmarking of stability tests. The phase limitation is applied to show that there is no direct counterpart in discrete time to the continuous-time off-axis circle criterion. Specificially, no Zames-Falb multiplier can be found to keep the stability if the phase limitation is reached. This is confirmed by numerical counterexample.

Phase limitations of both continuous-time and discrete-time Zames-Falb multipliers and their relation with the Kalman conjecture are analysed. A phase limitation for continuous-time multipliers given by Megretski is generalised and its applicability is clarified; its relation to the Kalman conjecture is  illustrated with a classical example from the literature.  It is demonstrated that there exist fourth-order plants where the existence of a suitable Zames-Falb multiplier can be discarded and for which simulations show unstable behavior. A novel phase-limitation for discrete-time Zames-Falb multipliers is developed. Its application is demonstrated with  a second-order counterexample to the Kalman conjecture. Finally, the discrete-time limitation is used to show that there can be no direct counterpart of the off-axis circle criterion in the discrete-time domain.

\end{abstract}

\section{Introduction}

The absolute stability of a negative feedback interconnection between an LTI system $G$ and a nonlinearity $\phi$ with a slope restriction $k$ has aroused the interests of many researchers. The stability tests include the circle criterion, Popov criterion \cite{Vidyasagar:1993}, \cite{Khalil:2002}, and off-axis circle criterion \cite{Cho:1968}, \cite{Harris:1983} in continuous time and the circle criterion \cite{Tsypkin:1962}, Tsypkin criterion \cite{Tsypkin:1964} and Jury-Lee criterion \cite{Jury:1964a}, \cite{Jury:1964b} in discrete time. For a recent discussion, see \cite{ahmad2013lmi} and \cite{ahmad2015less}. Apart from these, loop transformation and multiplier theory are both important tools to establish the stability of feedback interconnections. The Zames-Falb multipliers are a class of multipliers with the property of preserving the positivity of monotone and bounded nonlinearities, and hence of slope-restricted nonlinearities after loop transformation. The class of Zames-Falb multipliers can be defined in either continuous time \cite{OShea:1966, Zames-Falb:1968} or discrete time \cite{OShea:1967,Willems:1968}. 
Specifically, after loop transformation, the stability of the negative interconnection between an LTI system $G$ and a nonlinearity $\phi$ with a slope restriction $k$ is guaranteed if there exists a Zames-Falb multiplier $M$ such that
\begin{align}
\mbox{Re}\{M(1+kG)\}>0,
\end{align}
with $M$ and $G$ evaluated over all frequencies. That is to say, at $j \omega$, $\omega \in \md{R}$ for continuous-time systems and at $e^{j \omega}$, $\omega \in [0, 2\pi]$  for discrete-time systems.

The Zames-Falb multipliers may be considered a classical tool \cite{Desoer1975}. Nevertheless, there has been considerable recent interest, largely sparked by the availability of numerical searches (\cite{Safonov87},  \cite{chen1995robustness}, \cite{Chen95b}, \cite{Turner:2009}, \cite{Chang:2012}, \cite{Carrasco:2012}, \cite{Carrasco:2012c}, for continuous time; \cite{Syazreen:2013b}, \cite{Wang:2014} for discrete time) and their encapsulation within an IQC (integral quadratic constraint) framework \cite{megretski1997system}, \cite{kao2004matlab}, \cite{Veenman:2016}, \cite{Altshuller:2013}. There has also been interest in generalising the class, both to MIMO (multi-input, multi-output) nonlinearities \cite{safonov2000zames}, \cite{d2001new},  \cite{mancera2005all}, \cite{turner2009existence}, \cite{Fetzer:2017} and to nonlinearities outside the original classes considered by Zames and Falb \cite{rantzer2001}, \cite{materassi2011generalized},  \cite{Altshuller:2013},  \cite{Heath:2015b}. In addition to determining stability conditions, they can be used to analyse performance \cite{Turner:2011}, \cite{hu2016robustness}; further, they can be used to obtain tighter versions of the Popov criterion \cite{Carrasco:2013}. Applications of Zames-Falb multipliers range from input-constrained model predictive control \cite{heath07} to first order numerical optimisation algorithms \cite{lessard2016analysis}.

\begin{table}[t]
\caption{Various slope restrictions discussed in the text.}  %We must have $\hat{k}_{ZF} \leq k_{ZF} \leq k_S < \hat{k}_C \leq  k_N$. Similarly we must have $k_{ZF} < k_{PL}$. We conjecture that $k_{ZF}=k_S$.}
\label{table_slope_restrictions}
\begin{tabular}{|c|l|}\hline
$\hat{k}_{ZF}$ & Maximum slope for which a Zames-Falb multiplier is known \\ \hline
$k_{ZF}$ & Maximum slope for which there exists a Zames-Falb multiplier \\ \hline
$k_S$ & Maximum slope for which the Lur'e system is absolutely stable \\ \hline
$k_{PL}$ & Minimum slope for which phase limitation implies there is no\\
&  Zames-Falb multiplier \\ \hline
$\hat{k}_C$ & Minimum slope for which  a counterexample to absolute\\ 
&  stability is known\\ \hline
$k_{O}$ & Slope for direct discrete-time counterpart off-axis circle criterion \\
& (which is false) \\ \hline
$ k_{RO}$ & Slope for Reduced Off-axis circle criterion in \cite{Narendra:1968} \\ \hline
$k_N$ & Nyquist value \\ \hline
\end{tabular}
\end{table}

\begin{figure}[b]
  \centering
  \input{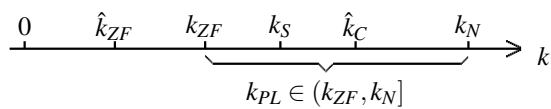}
  %\caption{Relationship of notations on the slope restrictions.}
\caption{Relations between slope restrictions discussed in the text. Conjecture~\ref{conj_carrasco} is that $k_{ZF}=k_S$ and hence $k_S < k_{PL}$.}
\label{k_axis}

\end{figure}

\begin{figure}[b]
  \centering
  \ifx\JPicScale\undefined\def\JPicScale{1}\fi
\unitlength \JPicScale mm
\begin{picture}(55,30)(0,0)
\linethickness{0.3mm}
\put(-1.25,25){\line(1,0){7.5}}
\put(6.25,25){\vector(1,0){0.12}}
\linethickness{0.3mm}
\put(8.75,5){\line(1,0){11.25}}
\linethickness{0.3mm}
\put(8.75,5){\line(0,1){17.5}}
\put(8.75,22.5){\vector(0,1){0.12}}
\linethickness{0.3mm}
\put(11.25,25){\line(1,0){8.75}}
\put(20,25){\vector(1,0){0.12}}
\linethickness{0.3mm}
\put(30,25){\line(1,0){12.5}}
\linethickness{0.3mm}
\put(42.5,7.5){\line(0,1){17.5}}
\put(42.5,7.5){\vector(0,-1){0.12}}
\linethickness{0.3mm}
\put(30,5){\line(1,0){10}}
\put(30,5){\vector(-1,0){0.12}}
\linethickness{0.3mm}
\put(42.5,5){\circle{5}}

\linethickness{0.3mm}
\put(45,5){\line(1,0){10}}
\put(45,5){\vector(-1,0){0.12}}
\put(50,7.5){\makebox(0,0)[cc]{$f$}}

\put(35,7.5){\makebox(0,0)[cc]{$v$}}

\put(15,28.75){\makebox(0,0)[cc]{$w$}}

\linethickness{0.3mm}
\put(20,30){\line(1,0){10}}
\put(20,20){\line(0,1){10}}
\put(30,20){\line(0,1){10}}
\put(20,20){\line(1,0){10}}
\linethickness{0.3mm}
\put(20,10){\line(1,0){10}}
\put(20,0){\line(0,1){10}}
\put(30,0){\line(0,1){10}}
\put(20,0){\line(1,0){10}}
\put(25,5){\makebox(0,0)[cc]{$\phi$}}

\put(25,25){\makebox(0,0)[cc]{$G$}}

\linethickness{0.3mm}
\put(8.75,25){\circle{5}}

\put(11.25,20){\makebox(0,0)[cc]{$-$}}

\put(1.25,28.75){\makebox(0,0)[cc]{$g$}}

\end{picture}
  \caption{Lur'e problem}\label{mp1}
\end{figure}
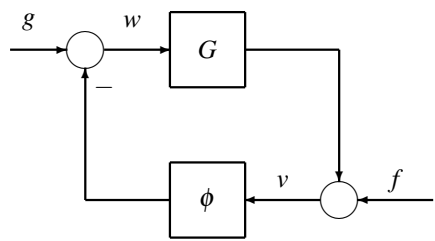

Although both continuous-time and discrete-time Zames-Falb multipliers are defined with similar conditions, there are clear distinctions between their properties. In discrete time the Zames-Falb multipliers are the full set of multipliers preserving the positivity of monotone and bounded nonlinearities, besides direct phase substitutions~\cite{Willems:1968},~\cite{Willems:1971}. In continuous time matters are more nuanced, but the class of Zames-Falb multipliers remains the widest known class of multipliers preserving the positivitiy of monotone and bounded nonlinearities, up to phase equivalence~\cite{Carrasco:2013, Carrasco:2014}. For a tutorial introduction to the phase properties of continuous-time Zames-Falb multipliers, phase-equivalence results and the issues associated with causality, see \cite{carrasco2016zames}. Phase properties  are essential to our understanding of Zames-Falb multipliers.
%Phase properties are key to the latter result, and they are essential to our understanding of Zames-Falb multipliers. %For example,  if $k<k_N$  (see Table~\ref{table_slope_restrictions} for various slope restrictions discussed in this paper), then the phase of $(1+kG)$ lies between $-180^\circ$ and $180^\circ$. Considerations of phase are crucial to the arguments of \cite{Carrasco:2012} and \cite{Carrasco:2013}. %In an early paper \cite{Freedman: 1972} the stability of the feedback is guaranteed by involving certain norms of the phase function of the plant $(1+kG)$ and its derivative. 

For example,  if $k<k_N$  (see Table~\ref{table_slope_restrictions} for various slope restrictions discussed in this paper), then the phase of $(1+kG)$ lies between $-180^\circ$ and $180^\circ$. Meanwhile multipliers must be positive so are restricted to lie between $-90^\circ$ and $90^\circ$. But as the Kalman conjecture is false, any set of suitable multipliers must be restricted by some further fundamental limitations. This follows from the obvious but important fact:
\begin{fact}\label{fact_ZF}
If  the system is not absolutely stable, there can be no appropriate Zames-Falb multiplier.
\end{fact}
However, only a few papers discuss such limitations. Megretski \cite{Megretski:1995} shows that there exists a phase limitation for continuous-time Zames-Falb multipliers. Another phase limitation of Zames-Falb multipliers is given by J{\"o}nsson and Laiou \cite{Jonsson:1996,Jonsson:PhD}. %Although that these limitations are given in the literature, they are numerically ill-conditioned, 
Such limitations are often ignored when new searches for multipliers are presented (see for example \cite{carrasco2016zames} and references therein). Often only $k_N$ is provided as an upper limit for the slope restriction.

We discuss Megretski's phase restriction \cite{Megretski:1995} with respect to a fourth-order continuous-time plant whose phase drops from $+180^\circ$  to $-180^\circ$; similarly with $k$ sufficiently big the phase of $1+kG$ drops from above $+90^\circ$ degrees to below $-90^\circ$. The limitation cannot be applied to first, second or third-order plants whose phase is in the range $(-180,90)$ degrees or $(-90,180)$ degrees; this agrees with the well-known result that the Kalman conjecture is true for such plants~\cite{Barabanov88}.  

To the best of authors' knowledge, no similar limitation has been developed in the discrete-time domain. Since there exist second-order discrete-time counterexamples to the Kalman conjecture whose phase is in the range $(-180,0)$ degrees~\cite{Carrasco:2015,Heath:2015} one might expect a \emph{simpler} limitation for discrete-time multipliers; this turns out to be indeed the case.% if the phase of $(1+kG)$ is outside the interval $(-90,90)$. 

The contribution of this paper is for both continuous-time and discrete-time multipliers.  We generalise Megretski's limitation \cite{Megretski:1995} for continuous-time multipliers to a wider choice of frequency intervals. Further, we show that Megrestki's limitation \cite{Megretski:1995} only applies for the class of Zames-Falb multipliers which do not require the odd condition on the nonlinearity; we provide the corresponding result when the nonliearity is odd.  We discuss  the limitation's numerical calculation and demonstrate its application in the context of  a classical example due to O'Shea~\cite{OShea:1966,carrasco2016zames}.  In particular we demonstrate a fourth-order counterexample of the Kalman conjecture for which the constraint is active.  A further contribution of the paper is the development of a phase limitation for discrete-time Zames-Falb multipliers. This limitation is fundamentally different to Megrestki's limitation as it only requires the phase of $(1+kG)$ to be either in the interval  $(90,180)$ degrees or  in the interval $(-180,-90)$ degrees. The limitation is easy to compute, and is active for a second-order discrete-time counterexample of the Kalman conjecture.
This close link  between the preclusion of a Zames-Falb multiplier and unstable behaviour leads us to the following conjecture as the counterpart to Fact~\ref{fact_ZF}; however no proof (or counterexample) is offered in this paper:

\begin{conj}\label{conj_carrasco}
If there is no appropriate Zames-Falb multiplier, the system is not absolutely stable.
\end{conj}

%However, the direct extension of the result in \cite{Megretski:1995} to discrete time is conservative. This is because frequency properties in continuous and discrete domain are fundamentally different. In continuous time the frequency is over the imaginary axis, but in discrete time it is over the unit circle. %Therefore the continuous time phase limitations cannot be applied to solve discrete-time systems. 
%Furthermore, the continuous-time Kalman conjecture is true for any first, second, and third-order system~\cite{Fitts:66,Barabanov88,Leonov:13} whereas the discrete-time Kalman conjecture is only true for any first-order system \cite{Carrasco:2015,Heath:2015}. %, since there exist second-order counterexamples~\cite{Carrasco:2015, Heath:2015}. 
%Therefore we might expect  a {\em simpler} phase restriction in discrete time. %In this paper we are looking for simpler criteria - i.e. phase limitations of discrete-time Zames-Falb multipliers, that may restrict the existence even for second-order discrete plant. Based on phase limitations, a slope restriction $k_p$ that is acting as a new benchmark of $k$ in multiplier search can be found, because it shows when $k = k_p$ there exists no discrete-time Zames-Falb multiplier such that the system is stable. 
%This is indeed the case, and is the main result of this paper.

One direct application of the phase limitation is to show there can be no direct discrete-time counterpart of the off-axis circle criterion. The continuous-time off-axis circle criterion is a useful graphical stability test and is shown to be a less conservative criterion compared to the circle criterion \cite{Cho:1968}, \cite{Harris:1983}. The derivation is based on the phase properties of RL/RC multipliers. %In \cite{Plummer:2000}, the discrete-time counterpart of the off-axis circle (described in Conjecture \ref{dis_off}) is proposed for  practical design. 
The direct discrete-time counterpart of the off-axis circle criterion is sometimes assumed to be true in the literature (e.g. \cite{Plummer:2000}, \cite{Okuyama:1999b}). However only a highly restrictive discrete-time version is proposed in \cite{Narendra:1968}, without discussion as to whether the direct discrete-time counterpart off-axis circle criterion is true or false. In this paper, we show that in some cases there are no Zames-Falb multipliers with the requisite phase properties for its derivation - i.e. the direct counterpart off-axis circle criterion cannot be derived using multiplier theory. The invalidation is completed by counterexample.% and explanation with phase limitations of Zames-Falb multipliers.

Some preliminary results related with Theorem IV.3 part~(i) were presented in \cite{Shuai:2015}.  %The structure of this paper is as follows. Some basic notations, definitions and background information are listed in Section II, including Megretski's continuous phase limitation and the off-axis circle criterion.  The main results of this paper are presented in Section III. We invalidate  the direct discrete-time counterpart off-axis circle criterion by interpretation in Section III.B. Numerical results are proposed to illustrate how to put the phase limitations into application in Section IV, with a counterexample of the direct discrete-time counterpart off-axis circle criterion. After analysis and discussion, conclusions and further work are given in Section V. 

\section{Notation and Preliminary results}

%Before moving onto the main results, some standard definitions and descriptions are listed in this section.

%\subsection{Definitions}

\subsection{Signal spaces}

For continuous-time signals let $\mathcal{L}_2[0,\infty)$ be the Hilbert space of square integrable and Lebesgue measurable functions $f:[0,\infty)\rightarrow\md{R}$ and let $\mathcal{L}_2$ be defined similarly for  $f:\md{R}\rightarrow\md{R}$.  Let $\mathcal{L}_{2e}[0,\infty)$ be the extended space of $\mathcal{L}_2[0,\infty)$ \cite{Willems:1971}.  %Let $\mathcal{L}_1$ be the space of all absolute integrable functions; given a function $h:\md{R}\rightarrow\md{R}$ such that $h\in\mathcal{L}_1$, its $\mathcal{L}_1$-norm is given by
%\begin{equation}
%\|h\|_1 = \int_{-\infty}^\infty |h(t)|\,dt.
%\end{equation}

For discrete-time signals let $\md{Z}$ and $\md{Z^+}$ be the set of integer numbers and positive integer numbers including $0$, respectively. Let $\ell$ be the space of all real-valued sequences, $h:\md{Z}^+\ra\md{R}$ and let $\ell_{2}$ denote the Hilbert space of all square-summable and measurable real sequences $f:\md{Z}^+\rightarrow\md{R}$  ($\ell$ is the extended space of $\ell_2$). Similarly, we can define the Hilbert space $\ell_2(\md{Z})$ by considering real sequences $f:\md{Z}\rightarrow\md{R}$. %Finally, let $\ell_1$ denote the Banach space of all absolute-summable real sequences. 
%Let $\ell_1$ be the space of all absolute summable real sequences; given a sequence $h:\md{Z}\rightarrow\md{R}$ such that $h\in\ell_1$, its $\ell_1$-norm is given by
%\begin{equation}
%\|h\|_1 = \sum_{k=-\infty}^\infty |h_k|.
%\end{equation}

\subsection{Lur'e problem and the Kalman conjecture}

The feedback interconnection system is a Lur'e system represented in Fig.~\ref{mp1} with both $G$ and $\phi$ mapping
$\mathcal{L}_{2e}[0,\infty)\rightarrow \mathcal{L}_{2e}[0,\infty)$ (continuous time) or $\ell\rightarrow\ell$ (discrete time). The object $G$ is assumed LTI stable and the object $\phi$ memoryless and slope-restricted (see below). The interconnection relationship is
\begin{equation}\label{eq:4}
	\begin{cases}
		v=f+Gw,  & \\
		w=-(\phi v)+g. & 
	\end{cases}
\end{equation}
%The discrete-time feedback interconnection system is represented in Fig.~\ref{mp1}. The nonlinearity $\phi$ is memoryless. The interconnection relationship is
%\begin{equation}\label{eq:4}
%	\begin{cases}
%		v(k)=f(k)+(Gw)(k),  & \\
%		w(k)=-(\phi v)(k)+g(k). & 
%	\end{cases}
%\end{equation}
The system~\eqref{eq:4} is well-posed if the map $(v,w)\mapsto(g,f)$ has a causal inverse on $\ell\times\ell$, and this feedback interconnection is $\ell_{2}$-stable if for any $f,g\in\ell_2$, both $w,v\in\ell_{2}$. 

%The nonlinear path of the Lur'e system belongs to the class of slope-restricted nonlinearities:

\begin{defn}
{\it (Memoryless slope-restricted nonlinearity)} The nonlinearity 
$\phi:\mathcal{L}_{2e}[0,\infty)\rightarrow \mathcal{L}_{2e}[0,\infty)$ or $\phi:\ell\rightarrow\ell$ is said to be memoryless and slope-restricted in $S[0,k]$, if there is a function $N:\md{R}\rightarrow\md{R}$ such that $(\phi u)(t)=N(u(t))$ or $(\phi u)(k)=N(u(k))$,  $N(0)=0$, and
		\begin{equation}\label{eq:7}
		    0\leq\frac{N (x_1)-N (x_2)}{x_1-x_2}\leq k, \quad \forall x_1, x_2\in\md{R}, x_1 \neq x_2.
		\end{equation} 
In addition, $\phi$ is said to be odd if $N$ is odd, i.e. $N(x)=-N(-x)$, for all $x\in\md{R}$.
\end{defn}

We define the Nyquist value and state the Kalman conjecture for both continuous-time and discrete-time systems. 

\begin{defn}[Nyquist value] Given a stable LTI system $G$, the Nyquist value $k_N$ is the supremum of all the positive real numbers $k$ such that $\tau k G$ satisfies the Nyquist Criterion for all $\tau\in[0,1]$. It can also be expressed as:
\begin{equation}
%    k_N=\sup\{k>0:\inf_\omega\{|1+\tau kG(e^{j\omega})|\}>0),\forall \tau \in[0,1]\}.
    k_N=\sup\{k>0:\inf_\omega\{|1+\tau kG|\}>0),\forall \tau \in[0,1]\},
\end{equation}
with $G$ evaluated over all frequencies (i.e. $\omega\in\mathbb{R}$ for continuous-time systems and $\omega\in[0,2\pi]$ for discrete-time systems).
\end{defn}

\begin{conj}[Kalman Conjecture, \cite{Kalman:1957}]
Let $\phi$ be a memoryless slope-restricted nonlinearity such that there exists a continuously differentiable $N: \md{R}  \rightarrow \md{R}$ and $S>0$ such that $\phi(v)(t) = N(v(t))$ (or $\phi(v)(k) = N(v(k))$) and 
\begin{equation}
    0 \leq \frac{d N(x)}{dx} \leq S, \quad \forall x \in \md{R}.
\end{equation}
Then the negative feedback interconnection of the continuous-time (or discrete-time) LTI systems $G \sim [A,B,C,0]$ and $\phi$ (Fig \ref{mp1}) is globally asymptotically stable if $A-BCk$ is Hurwitz (Schur) for all $k \in [0,S]$.
\end{conj}
There exist fourth-order continuous-time counterexamples to the Kalman conjecture \cite{Fitts:66,Barabanov88,Leonov:13} and second-order discrete-time counterexamples \cite{Carrasco:2015,Heath:2015}.

\subsection{Zames-Falb multipliers}

The characteristics of continuous-time Zames-Falb multipliers is given in the following theorem that defines two different classes of multipliers.

\begin{thm}\label{Thrm:1ct} 
\textbf{(Continuous-time Zames-Falb multipliers, \cite{Zames-Falb:1968}.)}
Consider the continuous-time feedback system in Fig.~\ref{mp1} with $G$ a stable LTI system  and $\phi$ memoryless and slope-restricted in $S[0,k]$. Suppose that there exists an LTI multiplier $M:\mathcal{L}_2\rightarrow\mathcal{L}_2$ whose transfer function has the form
\begin{equation}
	M(s)=1-H(s)
\end{equation}
such that the impulse response $h$ of $H$ satisfies
\begin{equation}
	\int_{-\infty}^\infty|h(t)|\,dt < 1.
\end{equation}
Moreover, let us assume that either $\phi$ is odd or $h(t)>0$. Suppose further there is some $\delta>0$ such that
\begin{equation}\label{ineq:6}
\mbox{Re}\left \{M(j\omega)(1+kG(j\omega)) \right \}\geq \delta\mbox{ for all }\omega\in\mathbb{R}.
\end{equation}
Then the feedback interconnection~\eqref{eq:4} is $\mathcal{L}_2$-stable.  \hspace*{\fill}~\IEEEQED\par\endtrivlist\unskip
\end{thm}
\begin{rem}
	With some abuse of notation, we denote $h(t)$ as the addition of a real-valued function $h_a(t)$ and impulses at different instants, i.e.
	\begin{equation}
	h(t)=h_a(t)+\sum_{i=1}^{\infty}h_i\delta(t_i).
	\end{equation}
\end{rem}

\begin{defn}
	The class of continuous-time Zames-Falb multipliers $\mathcal{M}^c$ is defined as the LTI systems $M:\mathcal{L}_2\rightarrow\mathcal{L}_2$ whose transfer function has the form
	\begin{equation}
	M(s)=1-H(s)
	\end{equation}
such that the impulse response $h$ of $H$ satisfies that $h(t)\geq0$ for all $t$ and
	\begin{equation}
	\int_{-\infty}^\infty h(t) \,dt< 1.
	\end{equation}
\end{defn}

\begin{defn}
	The class of continuous-time ``odd'' Zames-Falb multipliers $\mathcal{M}^c_{\text{odd}}$ is defined as the LTI systems $M:\mathcal{L}_2\rightarrow\mathcal{L}_2$ whose transfer function has the form
	\begin{equation}
	M(s)=1-H(s)
	\end{equation}
such that the impulse response $h$ of $H$ satisfies
	\begin{equation}
	\int_{-\infty}^\infty|h(t)| \,dt < 1.
	\end{equation}
\end{defn}
By definition, $\mathcal{M}^c\subset\mathcal{M}^c_{\text{odd}}$. 

The counterpart result in discrete time is given in the following theorem and it also defines two different classes of multipliers:

\begin{thm}\label{Thrm:1dt} 
\textbf{(Discrete-time Zames-Falb multipliers, \cite{Willems:1968}, \cite{Willems:1971})}
Consider the discrete-time feedback system in Fig.~\ref{mp1} with $G$ a stable LTI system and $\phi$ memoryless and slope-restricted in $S[0,k]$. Suppose that there exists an LTI multiplier $M:\ell_2\rightarrow\ell_2$ whose transfer function has the form
\begin{equation}
M(z)=1-H(z)
\end{equation}
such that the impulse response $h$ of $H$ satisfies that $h_0=0$ and 
\begin{equation}
\sum_{i=-\infty}^\infty |h_i|< 1.
\end{equation}
Moreover, let us assume that either the nonlinearity is odd or $h_i\geq0$. Suppose further
\begin{equation}\label{eq:14}
    \hbox{Re}\left\{M(z)(1+kG(z))\right\}>0, \quad \forall |z|=1.
\end{equation}
Then the feedback interconnection~\eqref{eq:4} is $\ell_2$-stable.  \hspace*{\fill}~\IEEEQED\par\endtrivlist\unskip
\end{thm}

\begin{rem}
Inequality (\ref{ineq:6}) is evaluated over $\omega\in\mathbb{R}$ whereas inequality (\ref{eq:14}) is evaluated over the frequency interval $\omega\in[0,2\pi]$. Hence, by the Extreme Value Theorem \cite{Apostol:1974}, it is unnecessary to define any $\delta>0$ for the discrete case corresponding to that used in the continuous case.
\end{rem}

Similarly to the previous definitions, we can define the classes of multipliers $\mathcal{M}^d$ and $\mathcal{M}^d_{\text{odd}}$.

\subsection{Off-axis circle criterion}

%For continuous-time systems, multipliers can be constructed in the form shown in Lemma~\ref{con_off_m}.
%
%{\color{blue} COMMENT: I think this Lemma is not required. It is a technicality that we use, and I have tried to explain later, but presenting this construction is misleading at this point.}
%
%\begin{lem}[Construction of continuous multipliers, \cite{Cho:1968}]\label{con_off_m}
%Let $[a,b]$ be any subinterval of $(0,\infty)$, $\theta$ be any constant in $(-\frac{\pi}{2},\frac{\pi}{2})$, $\epsilon >0$ be a constant and $M(j\omega)$ be a continuous multiplier that preserves the stability of the feedback interconnections.
%\begin{itemize}
%\item[1)] If $\theta \geq 0$, then there exists $M(j\omega)$ such that
%\begin{equation}
%|\angle M(j\omega)-\theta|<\epsilon, \forall \omega \in [a,b].
%\end{equation}
%\item[2)] If $\theta < 0$, then there exists $M(j\omega)$ such that
%\begin{equation}
%|\angle M(j\omega)-\theta|<\epsilon, \forall \omega \in [a,b].
%\end{equation}
%\end{itemize}
%\end{lem}

The continuous-time off-axis circle criterion is given. %based on Lemma \ref{con_off_m}.

\begin{lem}{\bf (Off-axis circle criterion for continuous-time\newline systems, \cite{Cho:1968})}\label{con_off}
Consider the feedback system in Fig.~\ref{mp1} with $G$ LTI stable and $\phi$ is slope-restricted in $S[0,k]$. Suppose that the Nyquist plot of the linear part of the system $G(j \omega)$ lies entirely to the right of a straight line passing through the point $(-\frac{1}{k}+\delta, 0)$ where $\delta >0$ and $\phi$ is monotonically increasing. Then the feedback interconnection~\eqref{eq:4} is $\mathcal{L}_2$-stable.
\end{lem}

%\begin{rem}
%In \cite{Cho:1968}, given the output $y(t)$ and internal state $x(t)$, the stability is defined in the sense that (1) $x(t) \in L_2$ implies $y(t) \in L_2$ and (2) $\lim_{t \rightarrow 0} |y(t)| = 0$. It is equivalent to $L_2$ stablity, i.e. $x(t) \in L_2$ implies $y(t) \in L_2$. In \cite{Narendra:2014}, it is shown that the off-axis circle criterion is only sufficient and not neccessary for the existence of an absolute Lyapunov function. 
%\end{rem}

For discrete time, only a highly restrictive version is proposed.

\begin{lem}{\bf (Reduced off-axis circle criterion for discrete-time systems, \cite{Narendra:1968})}\label{red_off}
Let the Nyquist plot of $G(e^{j\omega})$ for all $0 \leq \omega \leq \pi$ lie entirely to the right of a straight line, whose slope $k$ is nonnegative passing through $(-\frac{1}{K_2},0)$. Let $\omega_0$ be such that $\mbox{Re }G(e^{j\omega_0}) = -\frac{1}{K_2}$ and $\mbox{Re }G(e^{j\omega}) \geq -\frac{1}{K_2}$ for $\omega \geq \omega_0$ and $\mbox{Im }G(e^{j\omega}) \leq 0$ for $\omega_0 \geq \omega \geq 0$. Then the system is asymptotically stable for all monotone $\phi$ with slope restriction $K_2$ in the feedback path if
\begin{equation}
\theta \leq -\frac{1}{2} \omega_0 + \frac{\pi}{2},
\end{equation}
where $\theta$ is the angle made by the straight line and the imaginary axis, i.e., $\theta = \cot^{-1} k$. If $\mbox{Im }G(e^{j\omega}) \geq 0$ for $\omega_0 \geq \omega \geq 0$, the same argument can be used to prove the asymptotic stability of the system with nonpositive $k$ and
\begin{equation}
\theta \geq \frac{1}{2} \omega_0 - \frac{\pi}{2}.
\end{equation}
\end{lem}

\subsection{Further mathematical notation}

For the convenience of solving potential numerical issues, the notation of  $O(\cdot)$ is given. 

\begin{defn}
The condition
\begin{equation}
	f(t)=g(t)+O(t^n), \mbox{ as } t \rightarrow 0.
\end{equation}
means that there exist $M$ and $t_0$ such that
\begin{equation}
|f(t)-g(t)| \leq M t^n \mbox{ on }[0, t_0].
\end{equation}
\end{defn}

The floor function, denoted by $\lfloor \nu \rfloor $, is defined by
\begin{equation}
\lfloor \nu \rfloor =\max\{m\in \mathbb {Z} \mid m\leq  \nu\}.
\end{equation}

\section{Continuous phase limitations  and the Kalman conjecture}

Megretski presents in~\cite{Megretski:1995} a phase limitation for continuous-time Zames-Falb multipliers.
 In this section we generalise the result to a wider set of frequency intervals, and derive separate results for both $\mathcal{M}^c$ and $\mathcal{M}^c_{odd}$. Although it is stated in  \cite{Megretski:1995} that the result there is valid for $\mathcal{M}^c_{odd}$ (in the terminology of this paper) we show by counterexample that it is in fact valid for $\mathcal{M}^c$ only. 
%In this section, we revisit this phase limitation and clarify its validity, showing the result is only valid for $\mathcal{M}^c$. Moreover, we extend the result for the class of Zames-Falb multipliers  $\mathcal{M}^c_{\text{odd}}$. 
Finally, we bridge  the limitation of \cite{Megretski:1995} with the Kalman conjecture; this is the key motivation to develop a different  set of  phase limitations for the discrete-time Zames-Falb multipliers.

\subsection{Phase limitations}\label{sec:3a}

%\subsection{Megretski's phase limitation}
%In this section, we use a subclass of the continuous-time Zames-Falb multipliers given by
%\begin{equation}\label{def_CZF}
%M(s)=1-H(s)
%\end{equation}
%such that the inverse two sided Laplace transform of $H(s)$, $h(t)$, satisfies $h(t)\geq0$ and %$\int_{-\infty}^{\infty}h(t)dt\leq1$.  
\begin{figure}[t]
\centering
\includegraphics[width=0.45\textwidth]{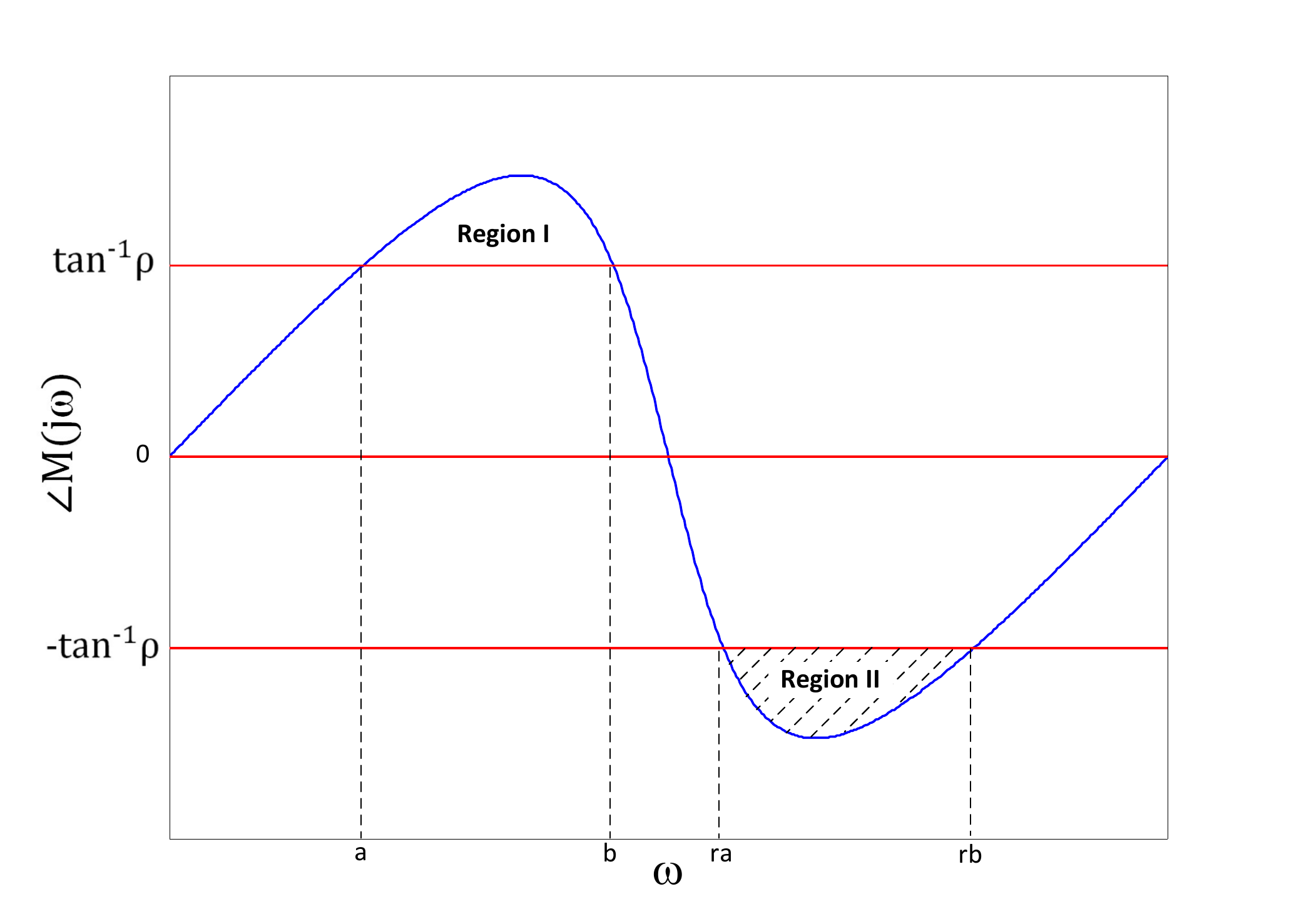}
%\vskip-15pt
\caption{\label{figure: physicalsense_con_bode} Illustration of Theorem~\ref{thm:limit_ct} with the choice of interval  from $\cite{Megretski:1995}$:  $\kappa=1$, $c=ra$ and $d=rb$. The result is given in terms of $\rho$ while the phase of the multiplier $M$ is $\tan^{-1} \rho$.} %{\color{red} Megretski's result (Theorem~\ref{Thrm:2}) is given in terms of $\rho$. The phase of the multiplier $M$ is $\tan^{-1} \rho$.}
%}}\vskip-15pt
\end{figure}

\begin{defn}\label{def:psi}
Let $0<a<b<c<d$,  $\kappa>0$, $\lambda > 0$ and $\mu>0$. Define
\begin{equation}\label{eq:Meg}
	\rho^c  =  \sup_{t>0} \frac{| \psi (t)|}{\phi (t)},
\end{equation}
and
\begin{equation}
	\rho^c_{odd}  =  \sup_{t>0} \frac{| \psi (t)|}{\tilde{\phi} (t)},
\end{equation}
where
	\begin{align}
\psi(t) & = \frac{\lambda \cos(at)}{t}  -\frac{\lambda \cos(bt)}{t}-\frac{\mu\cos(ct)}{t}+\frac{\mu\cos(dt)}{t},\\
\phi(t) & =  \lambda (b-a) +\kappa \mu (d-c) + \phi_1(t),
\end{align}
and
\begin{align}
\tilde{\phi}(t) & = \lambda (b-a) + \kappa\mu (d-c) - |\phi_1(t)|,
	\end{align}
with
\begin{equation}
\phi_1(t)  =  \frac{\lambda \sin(at)}{t}-\frac{\lambda \sin(bt)}{t}+ \frac{\kappa\mu\sin(ct)}{t}-\frac{\kappa\mu\sin(dt)}{t}.
\end{equation}
\end{defn}

\begin{lem}\label{lemma:psi}
%Define
%	\begin{align}
%\psi(t) & = \frac{\lambda \cos(at)}{t}  -\frac{\lambda \cos(bt)}{t}-\frac{\mu\cos(ct)}{t}+\frac{\mu\cos(dt)}{t},\\
%\phi(t) & =  \lambda (b-a) + \mu (d-c) + \phi_1(t),\\
%\tilde{\phi}(t) & = \lambda (b-a) + \mu (d-c) - |\phi_1(t)|,
%	\end{align}
%where
%\begin{equation}
%\phi_1(t)  =  \frac{\lambda \sin(at)}{t}-\frac{\lambda \sin(bt)}{t}+ \frac{\mu\sin(ct)}{t}-\frac{\mu\sin(dt)}{t},
%\end{equation}
%and where $d>c>b>a>0$, $\lambda >0$ and $\mu>0$.
If $\lambda$ and $\mu$ are chosen such that
\begin{equation}\label{eq:lambda_mu}
\frac{\lambda}{\mu} = \frac{d^2-c^2}{b^2-a^2},
\end{equation}
then $\rho^c$ and $\rho^c_{odd}$ in Definition~\ref{def:psi} are well-defined; that is to say $\rho^c<\infty$ and $\rho^c_{odd}<\infty$.
%we can define:
%\begin{enumerate}
%\item[(i)] $\rho^c<\infty$  as
%\begin{equation}\label{eq:Meg}
%	\rho^c  =  \sup_{t>0} \frac{| \psi (t)|}{\phi (t)};
%\end{equation}
%\item[(ii)]  $\rho^c_{odd}<\infty$ as
%\begin{equation}\label{eq:Meg}
%	\rho^c_{odd}  =  \sup_{t>0} \frac{| \psi (t)|}{\tilde{\phi} (t)}.
%\end{equation}
%\end{enumerate}
\end{lem}
%\emph{Proof:}
\begin{IEEEproof}
See Appendix.
\end{IEEEproof}
	
\begin{rem}
The direct calculation of the ratios $\psi(t)/\phi(t)$ and $\psi(t)/\tilde{\phi}(t)$ is numerically ill-conditioned for small~$t$ since, with the choice (\ref{eq:lambda_mu}), we have $\psi(t) = 0+O(t^3)$,  $\phi(t)=0+O(t^2)$ and $\tilde{\phi}(t)=0+O(t^2)$, all as $t\rightarrow 0$.
Nevertheless, the same construction ensures we can write
\begin{equation}
\frac{\psi(t)}{\phi(t)}  = \gamma t +O(t^3)\mbox{ and }
\frac{\psi(t)}{\tilde{\phi}(t)} = \gamma t +O(t^3)\mbox{ as }t\rightarrow 0,
\end{equation}
with
\begin{equation}
\gamma =  -\frac{1}{4}\frac{
\lambda(b^4-a^4)-\mu(d^4-c^4)
	}
	{
\lambda (b^3-a^3)+\kappa\mu (d^3-c^3)
	}.
\end{equation}
We use this relation for small $t$ in the numerical examples below.
\end{rem}

\begin{thm}[Continuous-time phase limitations]\label{thm:limit_ct}
Let $M$ be a continuous-time Zames-Falb multiplier. Suppose 
\begin{align}
	\mbox{Im} (M(j \omega)) &> \rho \mbox{Re} (M(j \omega))\mbox{ for all }\omega \in [a,b], \label{fre1}
\end{align}
and 
\begin{align}
	\mbox{Im} (M(j \omega)) &<- \kappa\rho \mbox{Re} (M(j \omega))\mbox{ for all }\omega \in [c,d], \label{fre2} 
\end{align}
for some $\rho>0$. Then under the conditions of Lemma~\ref{lemma:psi}.
\begin{enumerate}
\item[(i)] $\rho<\rho^c$ if $M\in\mathcal{M}^c$,
\item[(ii)] $\rho<\rho^c_{odd}$ if $M\in\mathcal{M}^c_{odd}$.
\end{enumerate}
\end{thm}

\begin{IEEEproof}
See Appendix.
\end{IEEEproof}

Lemma~\ref{lemma:psi} and Theorem~\ref{thm:limit_ct}, with the choice $\kappa=1$, $c=ra$, $d=rb$ and hence $\lambda/\mu=r^2$, are in \cite{Megretski:1995}. 
%\begin{thm}[Continuous phase limitation for $\mathcal{M}^c$, \cite{Megretski:1995}]\label{Thrm:2} 
%Let $b>a>0$, $r>b/a$ be real numbers. Let
%\begin{align}\label{eq:Meg}
%	\rho^c &= \rho^c(a,b,r) = \sup_{t>0} \frac{| \psi (t)|}{\phi (t)},
%\end{align}
%where
%\begin{equation}\label{psi_ct}
%\psi (t) =-\frac{\cos{(rbt)}}{rt}+\frac{r\cos{(bt)}}{t}-\frac{r\cos{(at)}}{t}+\frac{\cos{(rat)}}{rt}, 
%\end{equation}
%and
%\begin{multline}
%	\phi (t) = (r+1)(b-a) + \\ 
%+ \frac{ r\sin{(at)}}{t}- \frac{r\sin{(bt)}}{t}+ \frac{\sin{(rat)}}{rt}-\frac{\sin{(rbt)}}{rt}.
%\end{multline}
%Then $\rho^c < \infty$, and there exists no Zames-Falb multiplier \linebreak[4] $M\in\mathcal{M}^c$ such that 
%\begin{align}
%	\mbox{Im} (M(j \omega)) &> \rho \mbox{Re} (M(j \omega)), &w \in [a,b], \label{fre1}
%\end{align}
%and 
%\begin{align}
%	\mbox{Im} (M(j \omega)) &<- \rho \mbox{Re} (M(j \omega)), &w \in [ra,rb]. \label{fre2} 
%\end{align} \hspace*{\fill}~\QED\par\endtrivlist\unskip
%\end{thm}
An interpretation of Theorem~\ref{thm:limit_ct} with these values is illustrated in Fig.~\ref{figure: physicalsense_con_bode} (see also \cite{carrasco2016zames}). According to the constraints on the coefficients of continuous Zames-Falb multipliers,  if the phase is simultaneously greater than $\tan^{-1}\rho$ on $\omega \in [a, b]$ (in Region I) and smaller than $-\tan^{-1}\rho$ on $\omega \in[ra, rb]$ (in Region II), then $\rho<\rho^c$ if $M\in\mathcal{M}^c$ and $\rho<\rho^c_{odd}$ if $M\in\mathcal{M}^c_{odd}$. %If $rho$ is greater than these limits and the phase reaches Region I, then it will not reach Region II. 

\begin{rem} \label{rem:1} It is straightforward to produce the counterpart of Theorem~\ref{thm:limit_ct} with~\eqref{fre1} and~\eqref{fre2} replaced by $\mbox{Im}(M(j\omega))<-\rho \mbox{Re}(M(j\omega))$ for all $\omega\in[a,b]$ and  $\mbox{Im}(M(j\omega))>\kappa\rho \mbox{Re}(M(j\omega))$ for all $\omega\in[c,d]$ respectively.
\end{rem}

\begin{rem}
In \cite{Megretski:1995}, Megretski uses a positive sign in the exponential of the Laplace transform:
\begin{equation}
	M(j\omega)=1-\int_{-\infty}^{\infty}e^{j\omega t}h(t)dt.
\end{equation}
This is the standard convention in the Physics literature (see for example \cite{Bechhoefer2011}) but opposite to that used in \cite{Zames-Falb:1968}. The apparent discrepancy has no significant consequence for the analysis of phase limitations since if $M(s)$, with impulse response $m(t)$, is a Zames-Falb multiplier then $M(-s)$, with impulse response $m(-t)$, is also a Zames-Falb multiplier.
%%The phase limitation given in Theorem~\ref{Thrm:2} is working on the whole class of Zames-Falb multipliers. Actually, according to the proof, we found that it only works on the set of Zames-Falb multipliers with $h(t) > 0$. A first-order counterexample of the case $h(t)<0$ is given in the Appendix. 
\end{rem}

 It is natural to ask whether a phase limitation over a single frequency range can be constructed in a similar manner. This is not possible in continuous time, as any corresponding definition of $\rho^c$ or $\rho^c_{odd}$ would be unbounded as $t$ approaches~$0$. Loosely speaking, we can generate a multiplier in $\mathcal{M}^c$ with  phase arbitrarially close to $\pm90^\circ$ over an arbitrararily large frequency inteterval by selecting $h(t)=(1-\eps)\delta(t-t^*)$ with $\eps>0$ arbitrarially close to $0$ and $t^*$ arbitrarially close to 0. But we construct such phase limitations for discrete-time multipliers below, in Section~\ref{sec:dt}.

\subsection{Numerical example }
\begin{figure}[h]
	\centering
	\includegraphics[width=0.45\textwidth]{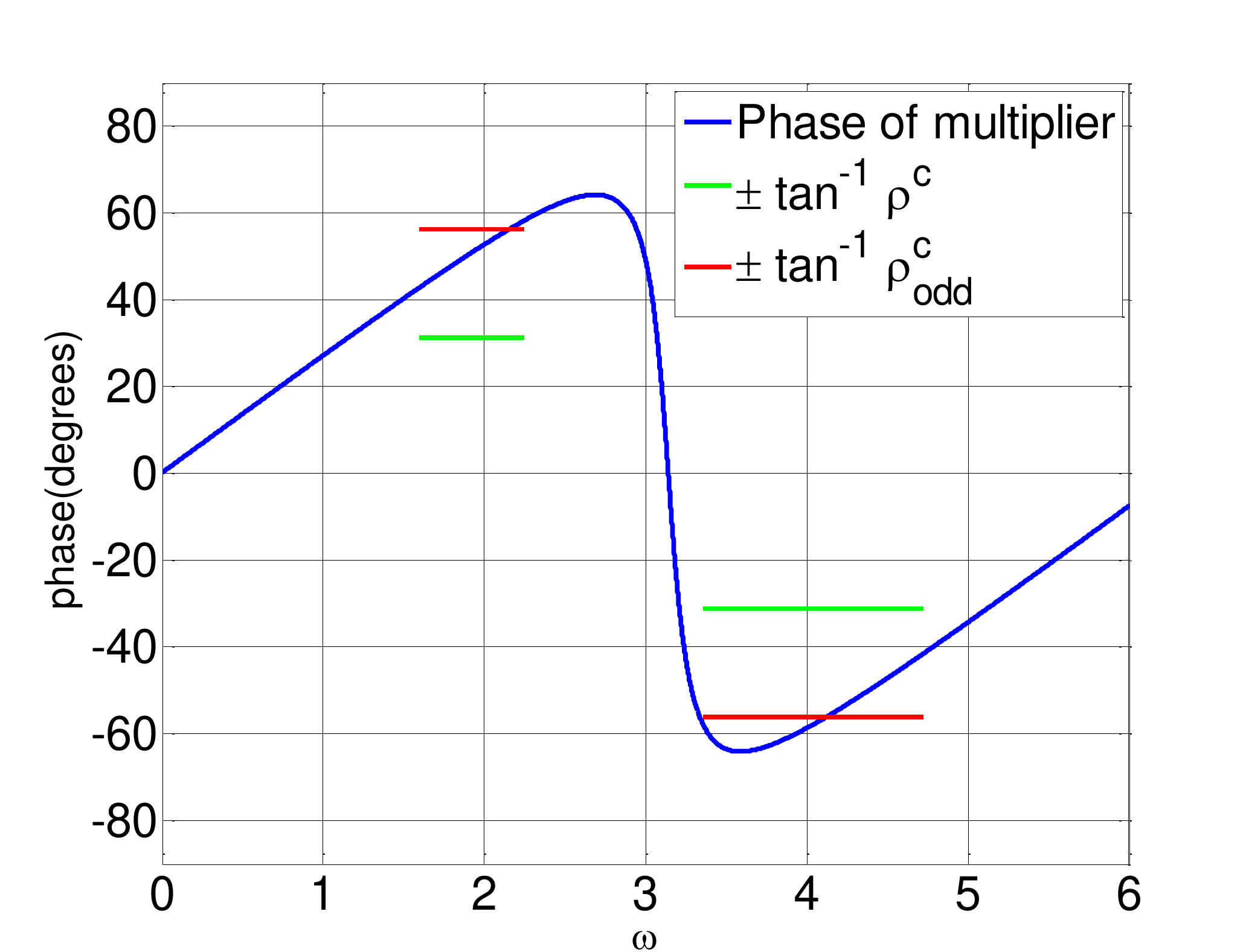}
	\caption{\label{figure: phase_counter_example}Phase of the multiplier (\ref{eq:multiplier}) and the continuous phase limitations $\pm \tan^{-1} \rho^c\approx \pm 31.25^\circ$ and $\pm \tan^{-1} \rho^c_{odd}\approx \pm 56.18$ evaluated on $[1.6,2.25]$ and $[3.36,4.725]$.}
\end{figure}

Here we illustrate Theorem~\ref{thm:limit_ct} with a numerical example. Let $a=1.6$ and $b=2.25$. Let   $\kappa=1$, $c=ra$ and $d=rb$ with $r=2.1$. Then a sweep over time intervals followed by local numerical search gives
\begin{equation}
\rho^c \approx 0.6069\mbox{, }\tan^{-1}\rho^c  \approx 31.25^\circ,
\end{equation}
and
\begin{equation}
\rho^c_{odd}  \approx 1.4928\mbox{, }\tan^{-1}\rho^c_{odd}  \approx 56.18^\circ.
\end{equation}
Now consider the multiplier
\begin{align}\label{eq:multiplier}
M(j \omega)= 1 - \int_{-\infty}^\infty  e^{-j \omega t}h(t)dt
\end{align} 
with $h(t)=-0.9 \delta(t+1)$. Figure~\ref{figure: phase_counter_example} shows that the relations (\ref{fre1}) and (\ref{fre2}), or equivalently
\begin{equation}
\angle M(j\omega) > \tan^{-1} \rho \mbox{ over the interval }[a,b],
\end{equation}
and
\begin{equation}
\angle M(j\omega) < -\tan^{-1} \rho \mbox{ over the interval }[ra,rb],
\end{equation}
are satisfied simultaneously for $\rho = \rho^c$ but not for $\rho = \rho^c_{odd}$. This is consistent with Theorem~\ref{thm:limit_ct} as $M\in\mathcal{M}^c_{odd}$ but $M\notin\mathcal{M}^c$. It is a counterexample to the false claim in \cite{Megretski:1995} that the phase limitation of Theorem~\ref{thm:limit_ct} part (i) is applicable to the wider class $M\in\mathcal{M}^c_{odd}$.

\begin{rem}\label{rem:sparse_ct}
Both in this numerical example and at the end of Section~\ref{sec:3a} we consider multipliers where $h$ takes the form $h(t)=(1-\varepsilon)\delta(t-\tau)$ for some $\tau$. There is a close link with Theorem~\ref{thm:limit_ct}. Specifically if $|\psi(\tau)|/\phi(\tau)=\rho^c$ and $\varepsilon\rightarrow 0$ then
%\begin{multline}
\begin{equation}
\int_{-\infty}^{\infty} \psi(t)h(t)dt 
=  \rho^c [\lambda (b-a) +\kappa\mu(d-c)]
+\rho^c \int_{-\infty}^{\infty}\phi_1h(t)\,dt.
\end{equation}
%\end{multline}
Compare (\ref{eq:ct_ineq1}) in the proof of Theorem~\ref{thm:limit_ct}. Similarly if $|\psi(\tau)|/\tilde{\phi}(\tau)=\rho^c_{odd}$ then
\begin{multline}
%\begin{equation}
\int_{-\infty}^{\infty} \psi(t)h(t)dt 
=   \\
=\rho^c_{odd} [\lambda (b-a) +\kappa\mu(d-c)]
+\rho^c_{odd} \int_{-\infty}^{\infty}\phi_1h(t)\,dt.
%\end{equation}
\end{multline}
We discuss the corresponding relations at greater length in Section~\ref{sec:4b} for the discrete-time case.
\end{rem}

%Briefly, we present an example with the different limits, and where a multiplier in $\mathcal{M}_{\text{odd}}$ does not satisfy the limitation in Theorem~\ref{Thrm:2}, but it satisfy the limitation in Theorem~\ref{Thrm:2a}. Let the phase limitation parameter $\rho$ in (\ref{eq:Meg}) be calculated over the intervals $[1.6, 2.25]$ and $[3.36, 4.725]$ by  setting $a=1.6$, $b=2.25$, and $r=2.1$. We sweep over $t$ from $t = 0.01$ and ending in $t=10$, with a step length of $0.001$. As a result, $\rho(1.6,2.25,2.1)<0.607$, resulting in a phase limitation of $31.26$ degrees. 

%We now show that it is possible to design a Zames-Falb multiplier with $h(t)<0$ that violates this constraint. Consider the Zames-Falb multiplier 
%\begin{align}
%M(j \omega)= 1 - \int_{-\infty}^\infty  e^{-j \omega t}h(t)dt
%\end{align} 
%with $h(t)=-0.99 \delta(t+1)$. The phase of this multiplier is clearly above $31.26$ degrees in the interval $[1.6, 2.25]$ and below $-31.26$ degrees in the interval $[3.36, 4.725]$ (see Fig~\ref{figure: phase_counter_example}). Therefore, this forms a counterexample.

%\subsection{Numerical example of Megrestki's limitation and our limitation}
%Briefly, we present an example with the different limits, and where a multiplier in $\mathcal{M}_{\text{odd}}$ does not satisfy the limitation in Theorem~\ref{Thrm:2}, but it satisfy the limitation in Theorem~\ref{Thrm:2a}. Let the phase limitation parameter $\rho$ in (\ref{eq:Meg}) be calculated over the intervals $[1.6, 2.25]$ and $[3.36, 4.725]$ by  setting $a=1.6$, $b=2.25$, and $r=2.1$. We sweep over $t$ from $t = 0.01$ and ending in $t=10$, with a step length of $0.001$. As a result, $\rho(1.6,2.25,2.1)<0.607$, resulting in a phase limitation of $31.26$ degrees. 

%We now show that it is possible to design a Zames-Falb multiplier with $h(t)<0$ that violates this constraint. Consider the Zames-Falb multiplier 
%\begin{align}
%M(j \omega)= 1 - \int_{-\infty}^\infty  e^{-j \omega t}h(t)dt
%\end{align} 
%with $h(t)=-0.99 \delta(t+1)$. The phase of this multiplier is clearly above $31.26$ degrees in the interval $[1.6, 2.25]$ and below $-31.26$ degrees in the interval $[3.36, 4.725]$ (see Fig~\ref{figure: phase_counter_example}). Therefore, this forms a counterexample.

%\begin{figure}[h]
%	\centering
%	\includegraphics[scale=0.63]{phase_counter_example2.pdf}
%	\caption{\label{figure: phase_counter_example}Phase of the multiplier $M(j \omega)$ and the continuous phase limitation}
%\end{figure}

\subsection{Counterexamples to Kalman conjecture via phase limitations}
It is instructive to interpret the phase limitations of Theorem~\ref{thm:limit_ct} in a manner consistent with known results about the Kalman conjecture. 

On the one hand, it is well-known that first, second and third order plants hold the Kalman conjecture~\cite{Barabanov88}. The phase of such plants cannot reach both Regions I and II in Fig.~\ref{figure: physicalsense_con_bode}. So  the phase limitations cannot apply to these plants. First-order plants do not require a dynamic multiplier, second-order plants require a dynamics multiplier with a tunable zero and a pole at infinity, i.e. a Popov multiplier, and-third order plant requires both a tunable pole and zero, i.e. first order RL/RC multipliers. In all these cases, only a first-order multiplier is required, and we know that there is no phase limitation in the selection of  such multipliers~\cite{Carrasco:2013}.
\begin{rem}
Although the off-axis circle criterion is also based on RL/RC multipliers, it is not sufficient to show all third-order plants hold the Kalman conjecture. %does not imply that it is possible to reach the Nyquist value of third order plants. 
For example, the off-axis circle criterion with the plant 
\begin{equation}
	G=\frac{s^2}{s^3+1.002s^2+s+0.998},
\end{equation}
guarantees stability with $k<3.928$, whereas the multiplier $M=(s+1)/(s+\eps)$ guarantees stability for any positive $k$ with a sufficiently small value $\eps>0$. 
\end{rem}

On the other hand the phase limitations may be applied to fourth-order plants, and these in turn may be counterexamples to the Kalman conjecture by: a) showing numerically that a phase limitation can be  applied to a well-known plant, and b) showing that the Lur'e system with this plant and a slope-restricted nonlinearity may be unstable. 

Specifically we will consider the phase limitation of Theorem~\ref{thm:limit_ct} part (i) with $\kappa=1$, $c=ra$, and $d=rb$; that is to say the original result of  \cite{Megretski:1995} applied to $\mathcal{M}^c$.
A particularly suitable example to show this limitation is O'Shea example \cite{OShea:1966,carrasco2016zames}:
\begin{equation}\label{plant}
G(s)=\frac{s^2}{(s^2+2\xi s+1)^2},
\end{equation}
since the symmetry of the problem simplifies the selection of the parameters. In this example, O'Shea showed that there is a Zames-Falb multiplier for any $k$ if $\xi>0.5$. The following result shows that it is not possible to reach an arbitrary large $k$ for any $\xi\leq0.25$. For the case $\xi=0.25$, the phase of $G(s)$ is above $177.98^\circ$  over the interval $[a,b]$ where $a=0.02249$ and $b=0.03511$; hence it is below $-177.98^\circ$  over the interval $[1/b,1/a]$ by using the symmetry of the plant. Then a suitable Zames-Falb multiplier for this plant would require a phase below $-87.98^\circ$ over the interval $[a,b]$ and above $87.98^\circ$ over the interval $[1/b,1/a]$. 
The phase limitation ensures that there is no Zames-Falb multiplier with such a phase characteristic, since $\tan^{-1}\rho^c\approx 87.79^\circ$. Strictly speaking, we have used the counterpart of Theorem~\ref{thm:limit_ct} mentioned in Remark~\ref{rem:1}.  
%The systematic use of Theorem~\ref{Thrm:2} to find the maximum value of $k$ is beyond the scope of this paper. 

\begin{figure}[h]
\centering
\includegraphics[trim=1.5cm 1cm 2cm 1cm,width=0.49\textwidth]{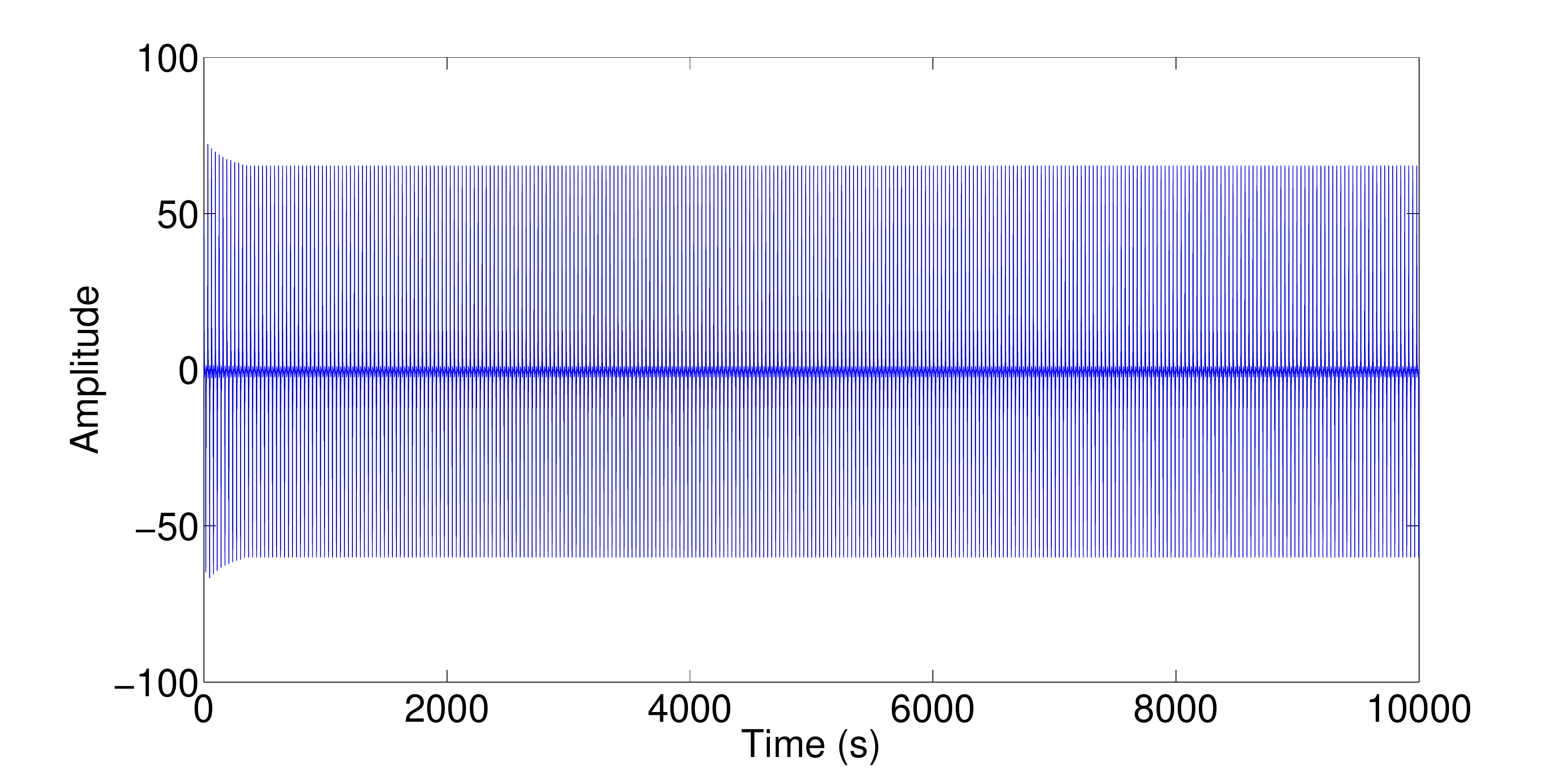}
%\includegraphics[width=5in]{fig5eps.eps}
%\vskip-10pt
\caption{\label{figure: uns}Amplitude of the signal $\nu$ in a simulation of the feedback interconnection depicted in Fig. 2 where G is given by~\eqref{plant} with $\xi=0.25$, $\phi$ is described by~\eqref{nonlinearity}, $g(t)$ is given by~\eqref{input}, and $f(t)=0$.}
%\vskip-10pt
\end{figure}

Although numerical reliability can be problematic in the discussion of the  Kalman conjecture~\cite{Leonov:13}, simulations of the plant with asymmetrical saturation show a time evolution that does not appear to settle to zero, supporting the validity of Conjecture~\ref{conj_carrasco}. The simulation shown in Figure~\ref{figure: uns} has been run in MATLAB R2013, using the solver ode45, with maximum step size of 0.0001 s, and relative tolerance of $10^{-3}$.  The nonlinearity $\phi$ is described by the nonlinear function
\begin{equation}\label{nonlinearity}
N(x)=\begin{cases}
-1000 & x<-1; \\
1000x     & -1\leq x\leq 0;\\
0   & x>0;
\end{cases} 
\end{equation}
the input $g$ is given by 
\begin{equation}\label{input}
g(t)=\begin{cases}
100 & t\leq 20 \text{s};\\
0     & t>20 \text{s};
\end{cases} 
\end{equation}
and $f(t)=0$.
The relevance of this counterexample to the Kalman conjecture is that we can show that there is no Zames-Falb multiplier with $h(t)\geq0$ for the system. The asymmetry of the nonlinearity seems to be a key factor as simulations with symmetric saturations show stable behaviour. The importance of asymmetry in the stability of Lur'e systems with saturation has been discussed recently \cite{Heath:2015b,Li:2016}.

\begin{rem}
The magnitude of the response in Fig.~\ref{figure: uns} is bounded. Since the plant~$G$ is stable and $\phi$ is sector-bounded it follows that if $g\in\mathcal{L}_{\infty}$ then all signals must be in $\mathcal{L}_{\infty}$.
\end{rem}

%The formal proof of Conjecture 1 is a challenging problem, but it would reduce the need to relay on simulations, since its validity can always be discussed due to its chaotic nature. Similarly it is challenging to produce non-conservative version of Theorem~\ref{Thrm:2} which can be numerically tested. A very elegant result is given in~\cite{Jonsson:PhD,Jonsson:1996}, but its numerical implementation is very difficult and it does not provide a straightforward interpretation in term of the phase of the multiplier. 
 
%This limitation indicates that there must be plants for which there exists no Zames-Falb multiplier such that the system is stable. By contrast, in discrete time there exist counterexamples to the Kalman conjecture for second-order plants~\cite{Carrasco:2015,Heath:2015}.
%This indicates that there may be more simple phase restrictions for discrete-time Zames-Falb multipliers, since the difference between maximum and minimum value of the desired multiplier phase for a second-order plant is within the range of $(0, 90^\circ)$ or $(-90^\circ,0)$ (see Figs. 7, 8, and 11).
% In discrete time, the counterexample of Kalman conjecture exists for second-order plant \cite{Carrasco:2015}. It means that, there exists a different phase limitation of discrete-time Zames-Falb multipliers. 

\section{Discrete-time phase limitation}\label{sec:dt}

In this section we develop  phase limitations for discrete-time Zames-Falb multipliers. Their derivation is in the spirit of Megretski's limitation \cite{Megretski:1995} and Theorem~\ref{thm:limit_ct} for continuous-time multipliers. However their properties are simpler and consistent with the existence of second-order discrete-time counterexamples to the Kalman conjecture~\cite{Carrasco:2015,Heath:2015}.  In particular, and by contrast with their continuous-time counterparts, they are concerned with properties over a single interval $\omega\in [a,b]$. 

%The previous section has provided a clear link between the Kalman conjecture and Megretski's phase limitation. Recent result on the discrete-time Kalman conjecture shows that it is possible to find second order counterexamples~\cite{Carrasco:2015,Heath:2015}. This suggest that phase-limitation in discrete-time Zames-Falb multiplier must be simpler as the phase of the multiplier must be limited as soon as it is required to be either close to $90$ or $-90$ degrees.

It is worth highlighting that if a discrete-time multiplier preserves  the positivity of all monotone and bounded nonlinearities then either it is  a Zames-Falb multiplier or there exists a Zames-Falb multiplier with the same phase~\cite{Willems:1968},~\cite{Willems:1971}. Hence any phase limitation on the discrete-time Zames-Falb multipliers is also a limitation for any discrete-time multiplier.

\subsection{Phase limitations}
\begin{defn}\label{def:psi_dt}
Let $0\leq a<b\leq \pi$. Define
\begin{equation}
	\rho^d  =  \max_{n\in\mathbb{Z}^+} \frac{| \psi_d (n)|}{\phi_d (n)},
\end{equation}
and
\begin{equation}
	\rho^d_{odd}  =  \max_{n\in\mathbb{Z}^+} \frac{| \psi_d (n)|}{\tilde{\phi}_d (n)},
\end{equation}
where
	\begin{align}
\psi_d(n)& = \frac{ \cos(an)}{n}  -\frac{\cos(bn)}{n},\\
\phi_d(n) & =   (b-a)  + \phi_{d,1}(n),
\end{align}
and
\begin{align}
\tilde{\phi}_d(n) & =  (b-a)  - |\phi_{d,1}(n)|,
	\end{align}
with
\begin{equation}
\phi_{d,1}(n)  =  \frac{ \sin(an)}{n}-\frac{ \sin(bn)}{n}.
\end{equation}
\end{defn}

\begin{lem}\label{lemma:psi_dt}
%Define
%	\begin{align}
%\psi(t) & = \frac{\lambda \cos(at)}{t}  -\frac{\lambda \cos(bt)}{t}-\frac{\mu\cos(ct)}{t}+\frac{\mu\cos(dt)}{t},\\
%\phi(t) & =  \lambda (b-a) + \mu (d-c) + \phi_1(t),\\
%\tilde{\phi}(t) & = \lambda (b-a) + \mu (d-c) - |\phi_1(t)|,
%	\end{align}
%where
%\begin{equation}
%\phi_1(t)  =  \frac{\lambda \sin(at)}{t}-\frac{\lambda \sin(bt)}{t}+ \frac{\mu\sin(ct)}{t}-\frac{\mu\sin(dt)}{t},
%\end{equation}
%and where $d>c>b>a>0$, $\lambda >0$ and $\mu>0$.
Both $\rho^d$ and $\rho^d_{odd}$ in Definition~\ref{def:psi_dt} are well-defined; that is to say $\rho^d<\infty$ and $\rho^d_{odd}<\infty$.
%we can define:
%\begin{enumerate}
%\item[(i)] $\rho^c<\infty$  as
%\begin{equation}\label{eq:Meg}
%	\rho^c  =  \sup_{t>0} \frac{| \psi (t)|}{\phi (t)};
%\end{equation}
%\item[(ii)]  $\rho^c_{odd}<\infty$ as
%\begin{equation}\label{eq:Meg}
%	\rho^c_{odd}  =  \sup_{t>0} \frac{| \psi (t)|}{\tilde{\phi} (t)}.
%\end{equation}
%\end{enumerate}
\end{lem}
\begin{IEEEproof}
See Appendix.
\end{IEEEproof}

\begin{thm}[Discrete-time phase limitations]\label{thm:limit_dt}
Let $M$ be a discrete-time Zames-Falb multiplier. Suppose 
\begin{align}
	\mbox{Im} (M(e^{j \omega})) &> \rho \mbox{Re} (M(e^{j \omega}))\mbox{ for all }\omega \in [a,b], \label{fre1_dt}
\end{align}
for some $\rho>0$. Then 
\begin{enumerate}
\item[(i)] $\rho<\rho^d$ if $M\in\mathcal{M}^d$,
\item[(ii)] $\rho<\rho^d_{odd}$ if $M\in\mathcal{M}^d_{odd}$.
\end{enumerate}
\end{thm}

\begin{IEEEproof}
	See Appendix.
\end{IEEEproof}

\begin{figure}[t]
\centering
\includegraphics[width=0.5\textwidth]{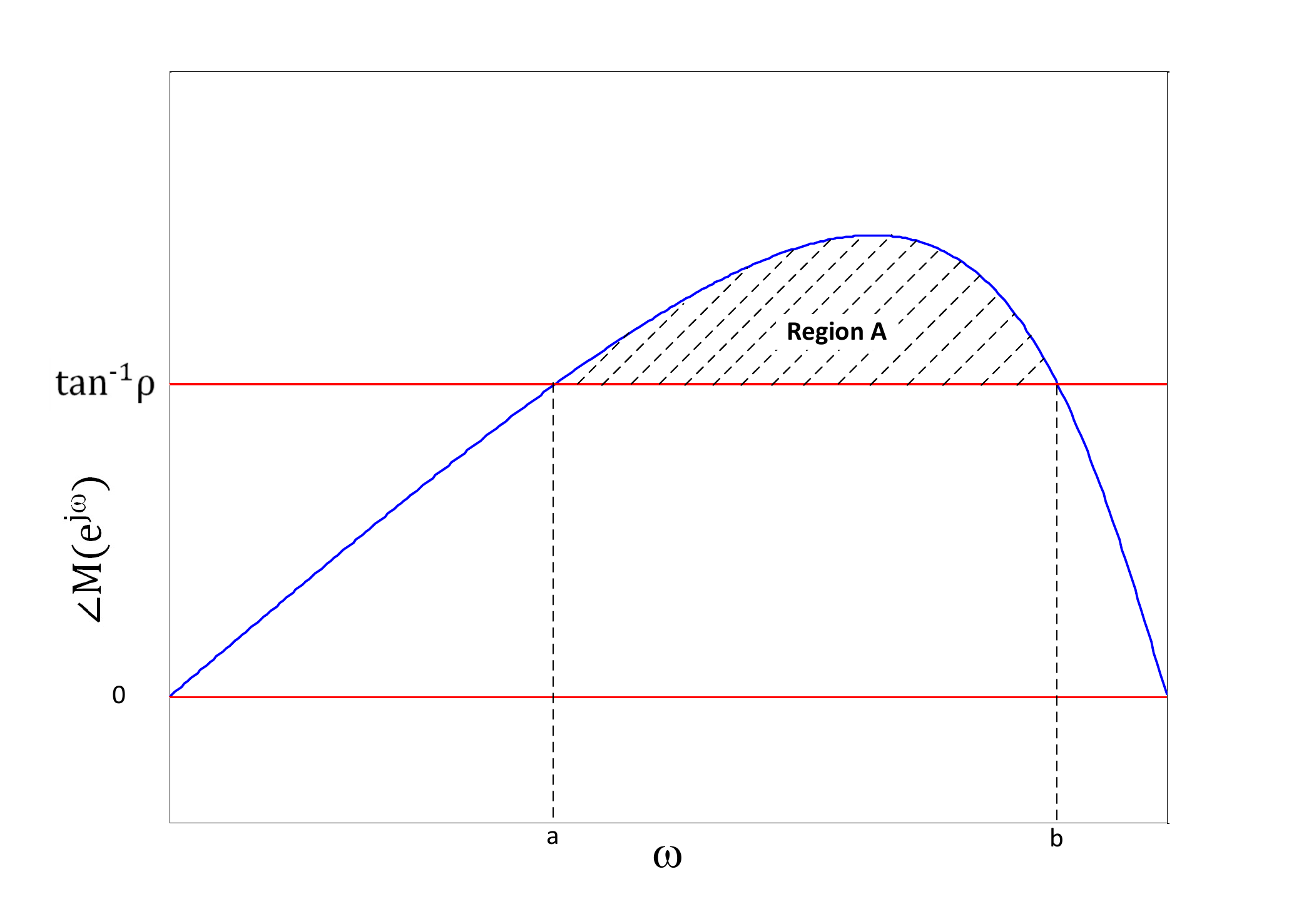}
\caption{\label{figure: physicalsense_dis_bode} %Discrete-time phase limitation result (Corollary~\ref{upper_lim}) is given in terms of $\rho_{\max}$. The phase of the multiplier $M$ is $\tan^{-1} \rho_{\max}$.
Illustration of Theorem~\ref{thm:limit_dt}. The result is given in terms of $\rho$ while the phase of the multiplier $M$ is $\tan^{-1}\rho$.}.
\end{figure}

An interpretation of Theorem~\ref{thm:limit_dt}  is illustrated in Fig.~\ref{figure: physicalsense_dis_bode}. According to the constraints on the coefficients of discrete-time Zames-Falb multipliers,  if the phase is greater than $\tan^{-1}\rho$ on $\omega \in [a, b]$ (in Region A), then $\rho<\rho^d$ if $M\in\mathcal{M}^d$ and $\rho<\rho^d_{odd}$ if $M\in\mathcal{M}^d_{odd}$. %If $rho$ is greater than these limits and the phase reaches Region I, then it will not reach Region II. 

\begin{rem} \label{rem:1dt} It is straightforward to produce the counterpart of Theorem~\ref{thm:limit_dt} with~\eqref{fre1_dt} replaced by 
\begin{equation}
\mbox{Im}(M(e^{j\omega}))<-\rho \mbox{Re}(M(e^{j\omega})) \mbox{ for all }\omega\in[a,b].
\end{equation}

\end{rem}
%An interpretation of Theorem~\ref{thm:limit_dt} is illustrated in Fig.~\ref{figure: physicalsense_dis_bode}. According to the constraints on the coefficients of discrete-time Zames-Falb multipliers, the phase cannot be  greater than $\tan^{-1}\rho_{\max}$ on the whole interval $ [a, b]$ (in Region A). Similarly, the phase cannot be smaller than $-\tan^{-1}\rho_{\max}$ on the whole interval $[a, b]$. 

An algorithm for finding the phase limitation in Theorem~\ref{thm:limit_dt} part (i) for a second order plant is given in~\cite{Shuai:2015}. For a given stable plant $G$ and a value of $k$ such that $0<k<k_N$ the phase of an ideal multiplier is obtained as
\begin{equation}\label{eq:phase}
\angle M_d=\begin{cases}
\angle (G+1/k)-90&  \text{if} \quad \angle (G+1/k)>90\\
\angle (G+1/k)+90& \text{if} \quad \angle (G+1/k)<-90\\
0&\text{otherwise}.\\
\end{cases}
\end{equation} 
Then the algorithm increases $k$ until the existence of such a multiplier  
can be discarded by using the limitation presented in Theorem~\ref{thm:limit_dt}. 

\subsection{Integral bound and sparsely parametrized multipliers}\label{sec:4b}
Theorem~\ref{thm:limit_dt}  gives relative bounds  on the real and imaginary parts  of a Zames-Falb multiplier's frequency response over an interval $[a,b]$. It is straightforward to derive a closely related result in terms of the integrals over the same interval.
\begin{thm}\label{thm:int_dt}
Let $M$ be a discrete-time Zames-Falb multiplier. Suppose 
\begin{align}\label{eq:int_dt}
	\int_a^b \mbox{Im} (M(e^{j \omega}))\, d\omega &> \rho \int_a^b \mbox{Re} (M(e^{j \omega}))\,d\omega
\end{align}
for some $\rho>0$. Then
\begin{enumerate}
\item[(i)] $\rho<\rho^d$ if $M\in\mathcal{M}^d$,
\item[(ii)] $\rho<\rho^d_{odd}$ if $M\in\mathcal{M}^d_{odd}$.
\end{enumerate}
\end{thm}

\begin{IEEEproof}
	See Appendix.
\end{IEEEproof}

\begin{rem}
Theorem~\ref{thm:int_dt} is stronger than Theorem~\ref{thm:limit_dt} in the sense that condition (\ref{fre1_dt}) is sufficient for condition (\ref{eq:int_dt}) but not necessary. Theorem~\ref{thm:limit_dt} may be derived as a Corollary of Theorem~\ref{thm:int_dt} by applying the Mean Value Theorem \cite{Apostol:1974}.
\end{rem}

Theorem~\ref{thm:int_dt} gives a tight phase limitation in the sense that we can associate a set of sparsely parameterized multipliers with Theorem~\ref{thm:int_dt} as follows.

\begin{prop}\label{thm:sparse}$\mbox{ }$
\begin{enumerate}
\item[(i)]For a given $a$ and $b$, define the set $\mathcal{N}^d\subset \mathbb{Z}$ as the set of integers $n$ such that $\psi_d(n)/\phi_d(n) =\rho^d$. Then multipliers of the form
\begin{equation}
		M(z) =  1- \sum_{n\in\mathcal{N}^d}h_n z^{-n}
\end{equation}
with
\begin{equation}
h_0=0\mbox{, }h_n \geq 0\mbox{ and }\sum_{n\in\mathcal{N}([a,b])}h_n=1-\varepsilon
\end{equation}
satisfy (\ref{eq:int_dt}) with $\rho$ arbitrarily close to $\rho^d$ in the limit as $\varepsilon\rightarrow 0$. 
\item[(ii)]For a given $a$ and $b$, define the set $\mathcal{N}^d_{odd}\subset \mathbb{Z}$ as the set of integers $n$ such that $\psi_d(n)/\tilde{\phi}_d(n) =\rho^d_{odd}$. Then multipliers of the form
\begin{equation}
		M(z) =  1- \sum_{n\in\mathcal{N}^d}h_n z^{-n}
\end{equation}
with
\begin{equation}
h_0=0\mbox{ and }\sum_{n\in\mathcal{N}([a,b])}h_n=1-\varepsilon
\end{equation}
satisfy (\ref{eq:int_dt}) with $\rho$ arbitrarily close to $\rho^d_{odd}$ in the limit as $\varepsilon\rightarrow 0$. 
\end{enumerate}
\end{prop}

\begin{IEEEproof}
	See Appendix.
\end{IEEEproof}

\begin{rem}  It is, once again, straightforward to produce the counterpart of Theorem~\ref{thm:int_dt} with~\eqref{eq:int_dt} replaced by 
\begin{equation}
\int_a^b\mbox{Im}(M(e^{j\omega}))\,d\omega<-\rho \int_a^b\mbox{Re}(M(e^{j\omega}))\,d\omega.
\end{equation} 
Similarly for Theorem~\ref{thm:sparse}  with (i) $\psi_d(n)/\phi_d(n)=-\rho^d$ and (ii) $\psi_d(n)/\tilde{\phi}_d(n)=-\rho^d_{odd}$.
\end{rem}

\begin{figure}[t]
\centering
\includegraphics[width=0.45\textwidth]{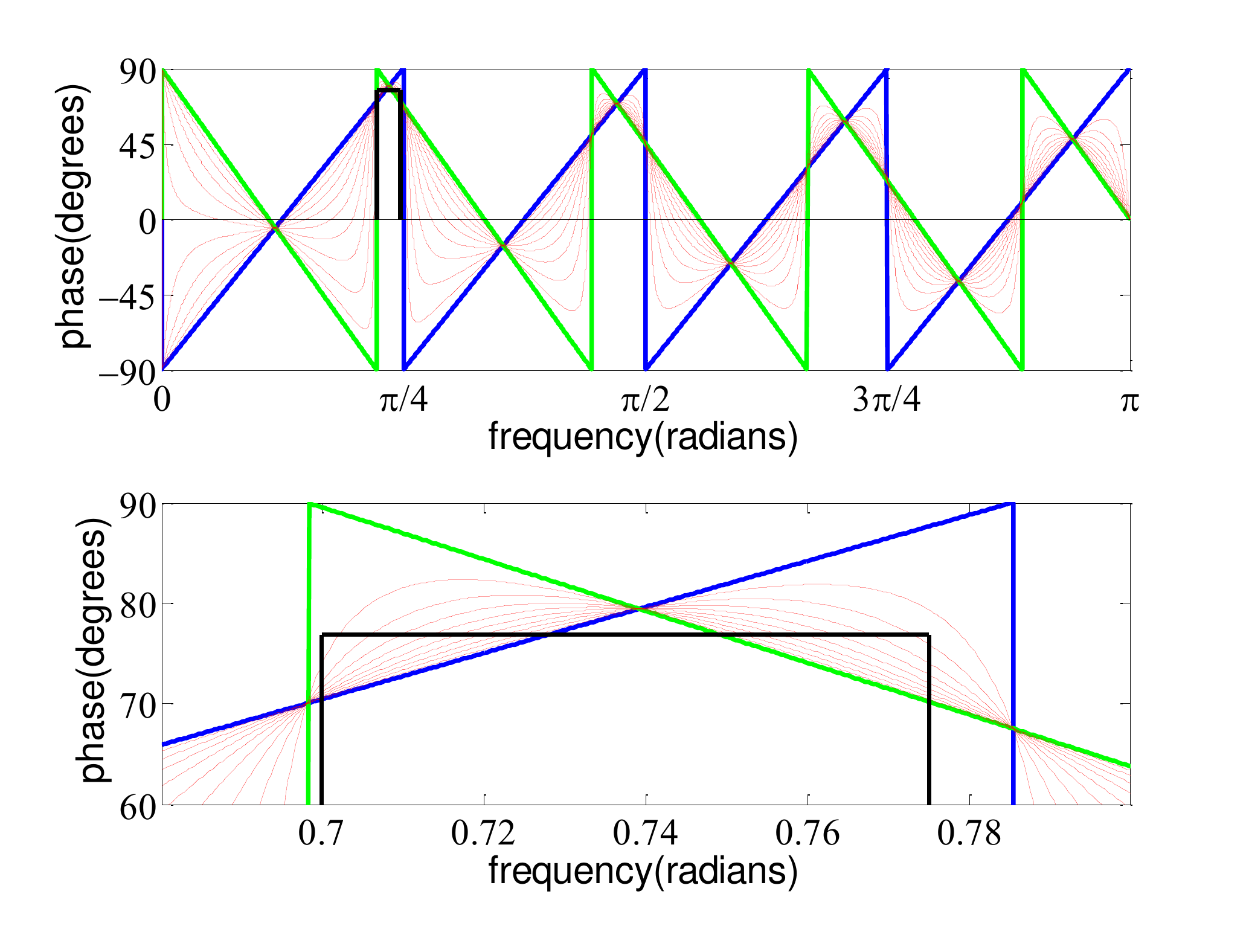}
\caption{\label{fig:limit_illustration}Phases of  the limiting cases  $M(z)=1-z^8$, $M(z)=1-z^{-9}$ and linear combinations of the form $M(z)=1-\lambda z^8 - (1-\lambda)z^{-9}$ with $0<\lambda<1$. The phase limitation $\tan^{-1} \rho^d \approx 76.8^\circ$ over the interval $[a,b]=[0.7,0.77501]$ is also shown. The top figure shows the phase over the frequency range from $0$ to $\pi$ radians, while the bottom figure shows the same data in the frequency range from $0.68$ to $0.8$ radians.}
\end{figure}

As an illustrative example, suppose $a=0.7$ and $b=0.77501$ (approx.). Then
\begin{equation}
\tan^{-1} \rho^d \approx 76.8^\circ,
\end{equation}
and
\begin{equation}
	\mathcal{N}^d = \left \{-8,9\right \}.
\end{equation}
Fig~\ref{fig:limit_illustration} shows the phases of the limiting cases  $M(z)=1-z^8$, $M(z)=1-z^{-9}$ and linear combinations of the form $M(z)=1-\lambda z^8 - (1-\lambda)z^{-9}$ with $0<\lambda<1$. It can be seen that the phases are near to $\tan^{-1} \rho^d$ over the interval $[a,b]$. However they always have values both above and below, indicating that Theorem~\ref{thm:limit_dt} is not tight in the same sense as Theorem~\ref{thm:int_dt}.

\begin{rem}
A similar analysis is possible for continuous-time multipliers. Compare Remark~\ref{rem:sparse_ct}.
\end{rem}

% It is easy to apply because %the ratio of the imaginary part and the real part of $M(j \omega)$ is related to $\angle M$ 
%\begin{equation}
%\angle M = \tan^{-1} \frac{\mbox{Im} (M(e^{j \omega}))}{\mbox{Re} (M(e^{j \omega}))}
%\end{equation}
%and this information can be obtained immediately from a Nyquist plot. %However, an algorithm for the non-conservative phase limitation is more challenging.

\subsection{Discrete-time counterparts of the off-axis circle criterion}

The off-axis circle criterion \cite{Cho:1968} (Theorem~\ref{con_off})  is a useful frequency-based graphical stability test for continuous-time systems. It is sometimes assumed (e.g. in~\cite{Plummer:2000})  that its discrete-time counterpart is true. We state this as a conjecture:

\begin{conj}\label{dis_off}
Consider the feedback system in Fig.~\ref{mp1} with $G\in\mathbf{RH}_\infty$, and $\phi$ is slope-restricted in $S[0,k]$. Suppose that the Nyquist plot of the linear part of the system $G(e^{j \omega})$ lies entirely to the right of a straight line passing through the point $(-\frac{1}{k}+\delta, 0)$ where $\delta >0$ and $\phi$ is monotonically increasing. Then the feedback interconnection~\eqref{eq:4} is $\ell_2$-stable.
\end{conj}

%\begin{rem}
%It is sometimes assumed (e.g. in~\cite{Plummer:2000} with an equivalent interpretation) that Conjecture~\ref{dis_off} is true. Apart from this, a reduced proposition and numerical examples are given in \cite{Narendra:1968} by implication.
%\end{rem}

A geometrical interpretation of both Theorem~\ref{con_off} for continuous-time systems and Conjecture~\ref{dis_off} for discrete-time systems is given in Fig \ref{geom_dis_oacc}. %Possible range of the phase of  the plant $(1+kG)$ with respect to the frequency $\omega$ is shown in Fig \ref{geom_dis_plant} in order to ensure the phase of $M(1+kG)$ to be within the range of $[-90^\circ, 90^\circ]$, as described in Fig \ref{geom_dis_plantM}.

%Conjecture~\ref{dis_off} is false. %Were it true, the Kalman conjecture would also be true for second-order discrete-time systems. Hence the counterexamples of \cite{Carrasco:2015} and \cite{Heath:2015} are also counterexamples of the Conjecture.  
%The results of this paper provide an interpretation in terms of phase of the differences between continuous-time, where the result is true~\cite{Cho:1968}, and discrete-time, where there can be no corresponding multiplier construction. We can summarise this argument as follows:
The phase-limitation on discrete-time Zames-Falb multipliers carries the implication that there can be no multiplier construction corresponding to that for RL/RC multipliers of \cite{Narendra:1968} used to prove Theorem~\ref{con_off}. We summarise the argument as follows:

\begin{figure}[h]
\centering
\includegraphics[width=0.45\textwidth]{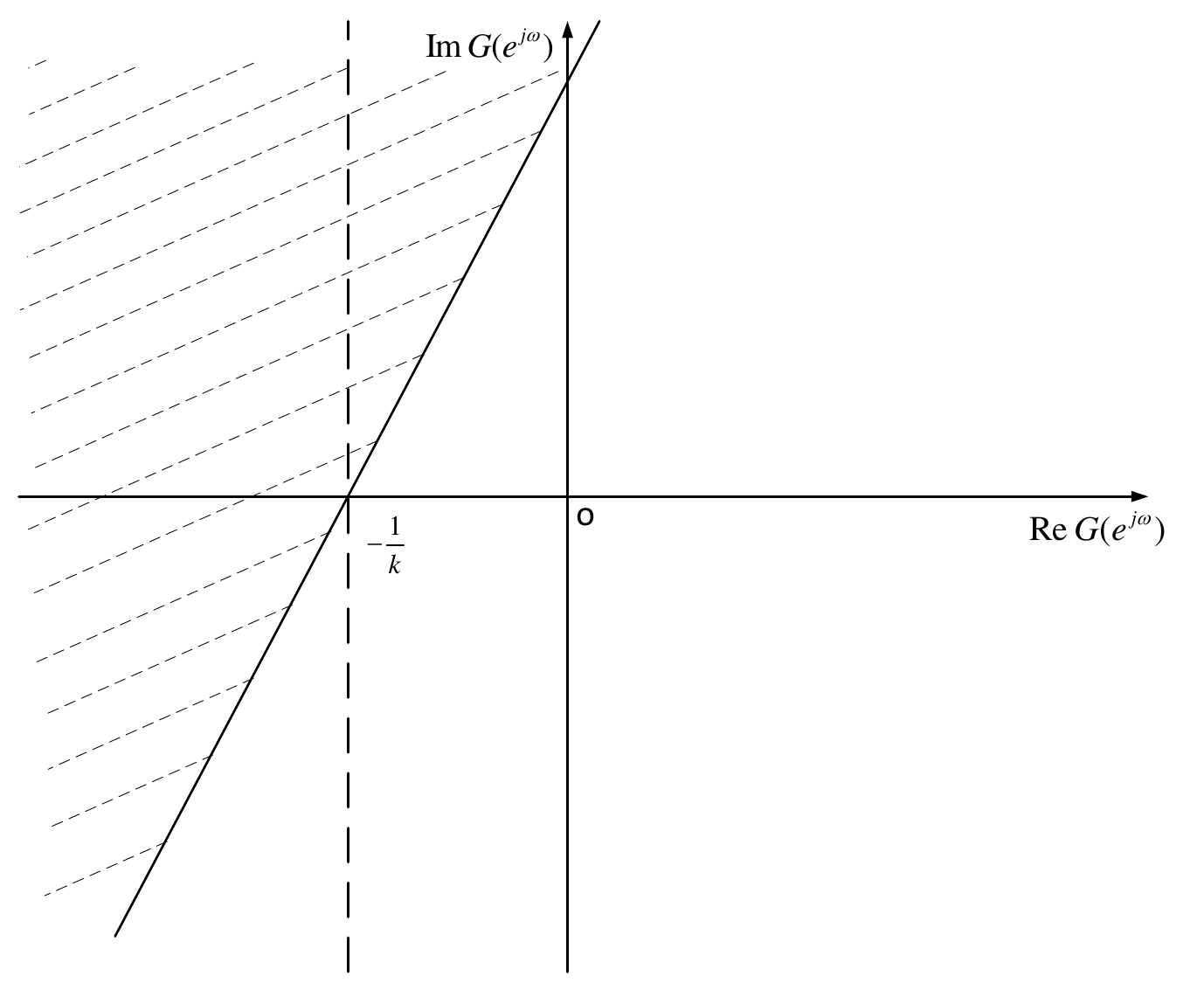}

\caption{Geometrical interpretation of the off-axis circle criterion considering the plant $G$ (Theorem~\ref{con_off} for continuous-time systems and Conjecture~\ref{dis_off} for discrete-time systems). The Theorem for continuous-time systems is true but the Conjecture for discrete-time systems is false in general.}\label{geom_dis_oacc}
\end{figure}

%\begin{figure}[h]
%\centering
%\includegraphics[scale=0.6]{geom_dis_plant.pdf}
%\vskip-4mm
%\caption{Possible range of the phase of  the plant $(1+kG)$ with respect to the frequency $\omega$}\label{geom_dis_plant}
%\end{figure}

%\begin{figure}[h]
%\centering
%\includegraphics[scale=0.6]{geom_dis_plantM.pdf}
%\vskip-4mm
%\caption{Possible range of the phase of $M(1+kG)$ with respect to the frequency $\omega$}\label{geom_dis_plantM}
%\end{figure}

\begin{enumerate}
\item Under the conditions of Conjecture \ref{dis_off} there is some $\theta$ in $(-90,90)$ degrees such that the phase of $1+kG$ always lies in the interval $(-90-\theta,90-\theta)$ degrees. Hence an ideal LTI multiplier with constant phase $\theta$ would render the real part of  $M(1+kG)$ positive over all frequencies. 
\item In their proof of the continuous off-axis circle criterion Cho and Narendra \cite{Cho:1968} show that it is possible to construct RL/RC multipliers whose phase is arbitrarily close to some constant $\theta$ degrees over an arbitrarily large interval. We show that for some values of $\theta$ this may not be possible for any discrete-time LTI multiplier. 
\item If a discrete-time LTI multiplier preserves the positivity of a slope-restricted nonlinearity then there is a Zames-Falb multiplier with the same phase~\cite{Willems:1968},~\cite{Willems:1971}, so we can limit our set of multipliers to the class of LTI Zames-Falb multipliers.
\item If $\theta>\tan^{-1}(2/\pi)\approx 32.48^\circ$ then Theorem~\ref{thm:limit_dt} precludes any such construction of a Zames-Falb multiplier since if $a\rightarrow 0^+$ and $b\rightarrow \pi^-$ then $\rho^d\rightarrow \tan^{-1}(-2/\pi)$. 
\end{enumerate}

%As a result, using a conservative criterion, we 
Hence the phase limitation can be used to invalidate Conjecture~\ref{dis_off} when $\theta>\tan^{-1}(2/\pi)$. Smaller values can be obtained by using different values of $a$ and $b$. 
It follows that any counterpart of the off-axis circle criterion in discrete-time must take into account specific information about frequency intervals. This is true of the more limited result originally derived by Narendra and Cho  \cite{Narendra:1968} (Theorem~\ref{red_off}).
In fact it can be shown that  the counterexamples  to the Kalman conjecture of \cite{Carrasco:2015} and \cite{Heath:2015} are also counterexamples to Conjecture~\ref{dis_off}.

\subsection{Finite search in discrete-time domain}

Here we provide a result which simplifies the numerical implementation. Although the definitions of $\rho^d$ and $\rho^d_{odd}$ are given with an infinite number of terms, they can be calculated using a finite number $n=n_N$ given in Lemma \ref{rho_bound_finite}. 

\begin{lem}\label{rho_bound_finite}
Let $0\leq a<b\leq \pi$, then
\begin{align}\label{rho_max1}
	\rho^d= \max_{1\leq n\leq n_N }\frac{| \psi_d(n)|}{\phi_d(n)},
\end{align}
and
\begin{align}\label{rho_max1odd}
	\rho^d_{odd}= \max_{1\leq n\leq n_N }\frac{| \psi_d(n)|}{\tilde{\phi}_d(n)},
\end{align}
with
\begin{align}
	n_N = \lfloor\nu\rfloor , \label{n_bound}
\end{align}
where
\begin{align}
	\nu =\frac{2(b-a)-2\sin b + 2\sin a -2 \cos b + 2 \cos a}{(a-b)(\cos b - \cos a)}.
\end{align}
\end{lem}

\begin{IEEEproof}
	See Appendix.
\end{IEEEproof}

\begin{figure}[h]
\centering
\includegraphics[width=0.45\textwidth]{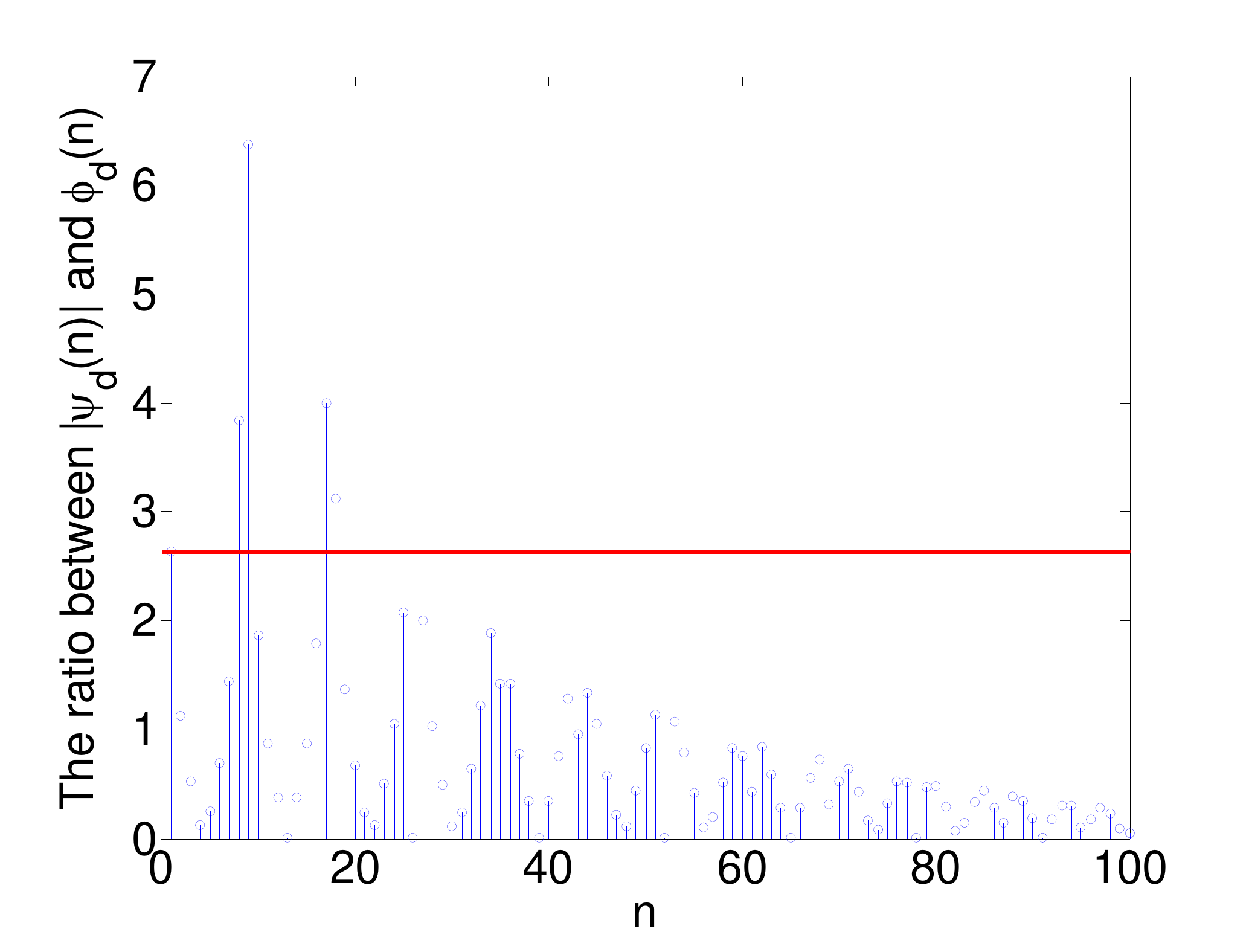}
%\vskip-10pt
\caption{\label{figure: zmlim_bound}The value of $f_d(n)$ with different value of $n$}
%\vskip-10pt
\end{figure}

Suppose we wish to find a phase limitation over the interval $\omega\in[0.7, 0.75]$. Applying Lemma~\ref{rho_bound_finite} we find $\nu =55.2$ and hence $|\psi_d(n)|/\phi_d(n) < |\psi_d(1)|/\phi_d(1)$ for all $n > n_N$, with
\begin{align}
	n_N = 55.
\end{align}
%It means $f(n)$ will hit the maximum value with $n^* \in [1, 56]$. To obtain $\rho_{\max}$, we can neglect $n> 56$ and only need to try finite times. 
Hence it is sufficient to search over the integers $1\leq n\leq 55$ for $\rho^d$. The numerical results shown in Fig. \ref{figure: zmlim_bound} demonstrate that $|\psi_d(n)| /\phi_d(n)< |\psi_d(1)|/\phi_d(1)$ for all $n > 18$. %As mentioned in Remark~\ref{Rem_opt}, {\color{red} since $\mathcal{N}=\{\pm 9\}$ and $f_d(9) < 0$ in this example, the Zames-Falb multiplier that reached higher (lower) values of phase over this interval is given by $M(z) = 1- r_{-9}z^{9}$ with $r_{-9} = 1- \epsilon$ ($M(z) = 1- r_9z^{-9}$ with $r_9= 1- \epsilon$), provided $\epsilon$ sufficiently small.
In fact the maximum  occurs at $n=-9$. %Theorem~\ref{upper_lim_con} says that (\ref{max_mean}) is satisfied with equality by $M(z)=1-(1-\varepsilon)z^9$ in the limit as $\varepsilon\rightarrow 0$. Similarly (\ref{max_mean_2}) is satisfied with equality by $M(z)=1-(1-\varepsilon)z^{-9}$ in the limit as $\varepsilon\rightarrow 0$.

%Once $n^*$ is obtained, the desired Zames-Falb multiplier within the interval $\omega \in [a, b]$ is in the form of $M(z) = 1- h_{n^*} z^{-n^*}$ according to Theorem \ref{lm_max}. Similarly, a non-causal Zames-Falb multiplier can be expressed as 

\subsection{Numerical example}

Let us consider the negative feedback interconnection between the plant
\begin{align}\label{example1}
G(z) = \frac{z}{z^2 - 1.8z +0.81},
\end{align}
and a slope-restricted nonlinearity. This second-order plant is a counterexample to the discrete-time Kalman conjecture as there is a periodic solution when $k = 2.1$ \cite{Carrasco:2015}, %Actually, our simulation results show that an oscillation for saturation can be found for $k_c = 1.7141$. 
and the Nyquist value is $k_N = 3.61$. Using the algorithm of \cite{Wang:2014} we find there exists a Zames-Falb multiplier for non-odd nonlinearities when $\hat{k}_{ZF}=1.3028$.

%Applying the algorithm of Section \ref{sec:alg}, we find $k_{PL}=1.4603$. The whole process is based on the standard line search of $k$. For a specific $k$, the existence of Zames-Falb multipliers is tested using phase limitation theorem. For example, when $k=1.5$, there exists no Zames-Falb multiplier such that the system is stable. The iterative testing procedure is illustrated in Fig. \ref{figure: algorithm1}. The data in each step is listed in Table \ref{table1}. We use a step length of $\epsilon = 0.1^\circ$. The desired phase $ -90^\circ  - \angle (1+kG(e^{j \omega}))$ reaches the maximum value when $\omega_0 = 0.8112$, which results in $\omega_1^{(0)} = \omega_2^{(0)} =0.8112$ and $-90^\circ  - \angle (1+kG(e^{j \omega_0})) = 67.7421^\circ$. When $i=1$, by solving $\angle (1+kG(e^{j 0.8112})) + \epsilon= \angle (1+kG(e^{j \omega}))$, we have $\omega_1^{(1)} =0.7656$ and $\omega_1^{(1)} =0.8560$. Then $\tan^{-1}\rho_{\max}^{(2)} = \tan^{-1}\rho_{\max}([0.7656, 0.8560])=78.0520^\circ$. Since $78.0520^\circ > 67.7421^\circ - \epsilon$, no phase limitation has been found. To increase the considered frequency period $[\omega_1, \omega_2]$, we move onto $i=2$ and we get $\omega_1^{(2)} =0.7466$ and $\omega_2^{(2)} =0.8742$. In this case,  $\tan^{-1}\rho_{\max}^{(3)} = \tan^{-1}\rho_{\max}([0.7466, 0.8742])=71.9596^\circ$. Keep doing this until $i=4$ and $\tan^{-1}\rho_{\max}^{(5)} =66.7137^\circ > 67.7421^\circ -4\epsilon$, where the phase limitation is reached. The phase limitation is given as the lowest red line in Fig. \ref{figure: algorithm1}. 

\begin{figure}[h]
\centering
\includegraphics[width=0.5\textwidth]{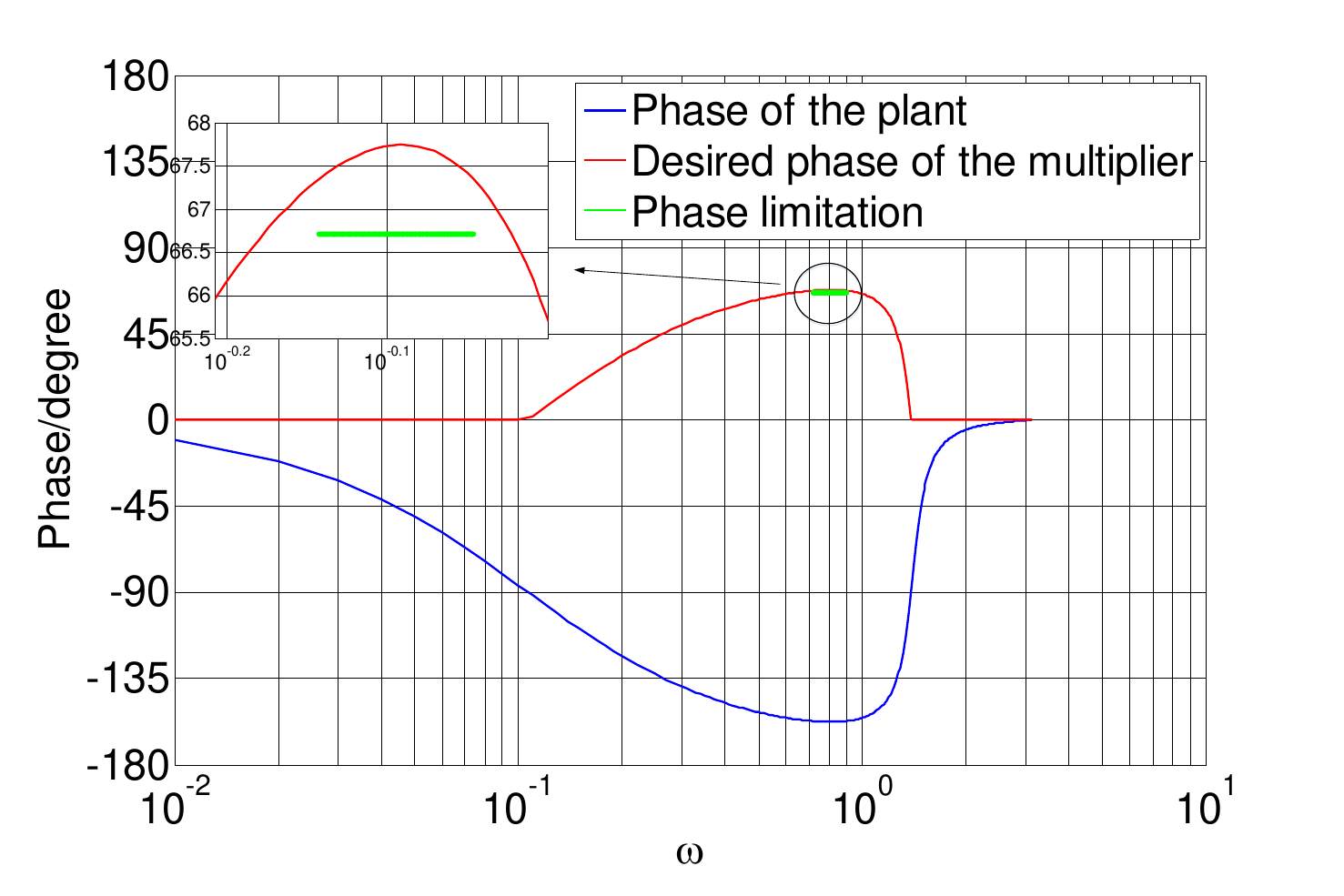}
%\vskip-15pt
\caption{\label{figure: algorithm_stepbystep_new3}Phase of $(1+1.5G)$, desired phase of the multiplier and the phase limitation}
%\vskip-14pt
\end{figure}

\begin{figure}[h]
\centering
\includegraphics[width=0.5\textwidth]{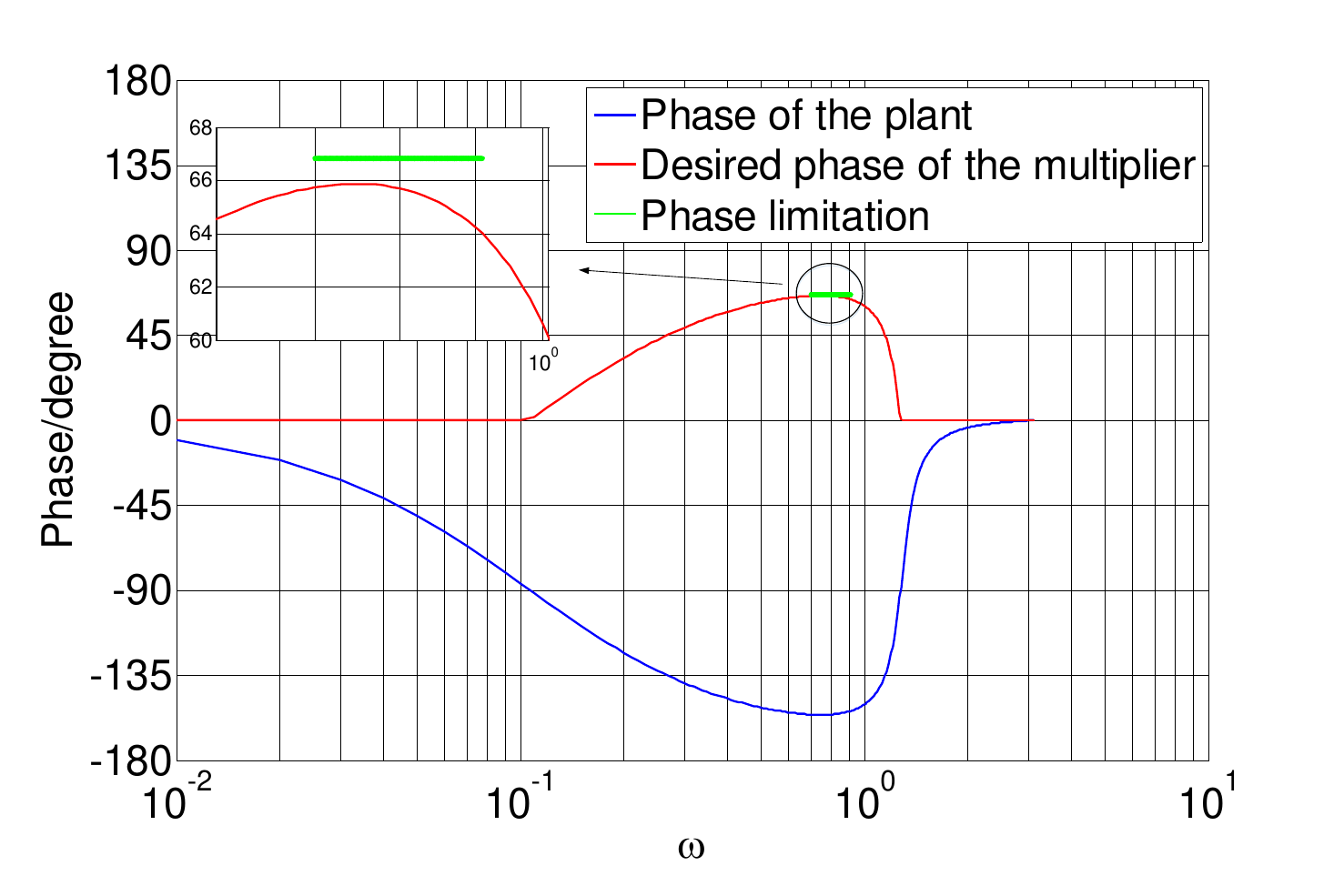}
%\vskip-15pt
\caption{\label{figure: phase_kZF2}Phase of $(1+1.3028G)$, desired phase of the multiplier and the phase limitation}
%\vskip-14pt
\end{figure}

Using the phase limitation result given in Theorem~\ref{thm:limit_dt} part~(i), it is possible to show that there is no Zames-Falb multiplier for any $k>k_{PL}=1.4603$. %Numerical results on the nonexistence of a Zames-Falb multiplier due to this phase limitation is given in Fig. \ref{figure: algorithm_stepbystep_new3} for $k=1.5$. {\color{red} The phase limitation is given by $\tan^{-1}\rho_{\max}=66.7137^\circ$, where $\rho_{\max}$ is obtained using Definition~\ref{rho_def} with  $a = 0.7198$ and $b=0.8996$. By contrast, Fig. \ref{figure: phase_kZF2} shows that this limitation is not active when $k=\hat{k}_{ZF}$, as expected since we have been able to find a suitable Zames-Falb multiplier for this value of the gain. The result of reduced off-axis circle criterion $k_R$ shows conservativeness compare to all the other results in the Table. Result from direct discrete-time counterpart off-axis circle criterion is greater than the slope obtained by phase limitation, i.e. $k_O > k_{PL}$ demonstrated that Conjecture~\ref{dis_off} is false.}
Fig. \ref{figure: algorithm_stepbystep_new3} illustrates that the phase limitation results indicate there can be no appropriate Zames-Falb multiplier when  $k=1.5$. The phase limitation is given by $\tan^{-1}\rho^d=66.7137^\circ$, where $\rho^d$ is obtained using Definition~\ref{def:psi_dt} with  $a = 0.7198$ and $b=0.8996$. By contrast, Fig. \ref{figure: phase_kZF2} shows that this limitation is not active when $k=\hat{k}_{ZF}$; this is expected since we have been able to find a suitable Zames-Falb multiplier for this value of the gain. A complete list of slope restriction results of $G(z)$ in (\ref{example1}) is given in Table~\ref{slope_restrictions_exa1}. The result of the reduced off-axis circle criterion $k_R$ shows conservativeness compared to all the other results in the Table. The (false) result from the direct discrete-time counterpart  of the off-axis circle criterion is greater than the slope obtained by phase limitation, i.e. $k_O > k_{PL}$; this  demonstrates that Conjecture~\ref{dis_off} is false.

Finally, using combination of deadzone and saturation as nonlinearity, we are able to find periodic solution with $\hat{k}_C=1.3666$. These results are consistent with Conjecture~\ref{conj_carrasco}, i.e. $\hat{k}_{ZF}<\hat{k}_C<k_{PL}$. 

%A complete list of slope restriction results of $G(z)$ in (\ref{example1}) is given in Table \ref{slope_restrictions_exa1}.

\begin{table}[h]
\centering
\caption{Results of different slope restrictions (non-odd nonlinearity)}
\label{slope_restrictions_exa1}
\begin{tabular}{|c|c|c|c|c|c|c|}
\hline
$k$     & $k_{RO}$ & $\hat k_{ZF}$ & $\hat k_C$ & $k_{PL}$ &  $k_O$ & $k_N$ \\ \hline
Result  & 0.8962       &1.3028 & 1.3666 & 1.4603       &   3.61  & 3.61  \\ \hline
\end{tabular}
\end{table}

\section{Conclusions}

In this paper we have demonstrated the connection between phase limitations of Zames-Falb multipliers and the Kalman conjecture. 

In continuous time, we have generalised a limitation proposed by Megretski, clarified its remit and  illustrated its effect with a numerical example. %A numerical result is require to obtain numerical well-conditioned values for the limitation. 
In particular we show it can be applied to a fourth-order plant where the resulting numerical implementation shows instability. % addition, we show that the Lur'e system where the limitation ensure that there is no suitable Zames-Falb multipleir is unstable.
It remains open which choice of intervals $[a,b]$ and $[c,d]$ and scaling parameter~$\kappa$ in Theorem~\ref{thm:limit_ct} provides most insight.

Motivated by this connection and recent results on the Kalman conjecture in discrete time, we have derived a more simple phase limitation for discrete-time Zames-Falb multipliers. Numerical results in discrete time are easier to obtain and we show that the slope restriction obtained by using phase limitation theorems can be about 40\% of the Nyquist value even for some second-order examples. Thus the phase limitation can be directly useful when forming  benchmarks for searches over Zames-Falb multipliers.  %As an appliccation of the phase limitation, the direct discrete-time counterpart off-axis circle criterion is invalidated, by both interpretation and numerical counterexample. 
Further, the phase limitation can be used to show there can be no direct counterpart in discrete time (Conjecture \ref{dis_off}) to the off-axis circle criterion for continuous-time systems (Theorem \ref{con_off}).%
%If the direct counterpart of the off-axis circle criterion (Conjecture \ref{dis_off}) has already been invalidated via numerical results, the developed limitation allows us to provide some insights on this conjecture and the further conditions introduced by Narendra and Cho in their discrete-time counterpart of the off-axis circle criterion.

Based on the results of this paper, we propose Conjecture~\ref{conj_carrasco}, which seems to be compatible with current state-of-the-art knowledge and results for both continuous and discrete-time domains. 

There is plenty of scope for future work. %In particular, we do not consider the wider class of Zames-Falb multipliers available when the nonlinearity is odd. 
It seems possible that the phase limitations might be used to provide a more computationally efficient search fo appropriate multipliers.
Phase limitations for the class of Zames-Falb multipliers available when the nonlinearity is quasi-odd \cite{Heath:2015b} require further research. 

\section{Acknowledgements}
We would like to thank the anonymous reviewers for their helpful suggestions.

\section{Appendix}

\subsection{Proof of Lemma~\ref{lemma:psi}}
Both the functions 
\begin{equation}
f_1(\omega) = \omega - \frac{\sin \omega t}{t}
\end{equation}
and 
\begin{equation}
f_2(\omega) = \omega + \frac{\sin \omega t}{t}
\end{equation}
are monotone non-decreasing in $\omega$ when $t>0$. It follows that $\phi(t)>0$ and $\tilde{\phi}(t)>0$ when $t>0$. In addition
\begin{equation}
\lim_{t\rightarrow \infty} \phi(t)  = \lambda(b-a)+\kappa\mu(d-c)>0,
\end{equation}
and
\begin{equation}
\lim_{t\rightarrow \infty} \tilde{\phi}(t) = \lambda(b-a)+\kappa\mu(d-c)>0.
\end{equation}
Finally
\begin{multline}
\phi_1(t)  =   -[\lambda (b-a) + \kappa\mu (d-c)]\\
+ \lambda \frac{(b^3-a^3)t^2}{6}+\kappa\mu\frac{(d^3-c^3)t^2}{6}+O(t^4)\mbox{ as }t\rightarrow 0,
\end{multline}
and
\begin{multline}
\psi(t)  =  \lambda\frac{(b^2-a^2)t}{2}-\mu\frac{(d^2-c^2)t}{2}\\
-\lambda\frac{(b^4-a^4)t^3}{24}+\mu\frac{(d^4-c^4)t^3}{24}+O(t^5)\mbox{ as }t\rightarrow 0,
\end{multline}
so the choice (\ref{eq:lambda_mu}) ensures 
\begin{equation}
\lim_{t\rightarrow 0} \frac{|\psi(t)|}{\phi(t)} = 0,
\end{equation}
and
\begin{equation}
\lim_{t\rightarrow 0} \frac{|\psi(t)|}{\tilde{\phi}(t)} = 0.
\end{equation}

\subsection{Proof of Theorem~\ref{thm:limit_ct}}

Suppose (\ref{fre1}) and (\ref{fre2}) hold for some multiplier $M(j\omega)=1-H(j\omega)$.
Then
\begin{equation}
\mbox{Im}(M(j\omega))  = \int_{-\infty}^{\infty}\sin(\omega t) h(t)\,dt,
\end{equation}
and
\begin{equation}
\mbox{Re}(M(j\omega))  =1 -  \int_{-\infty}^{\infty}\cos(\omega t) h(t)\,dt,
\end{equation}
where $h$ in the impulse response of $H$.
Hence integrating (\ref{fre1}) and (\ref{fre2}) over their respective intervals gives
\begin{multline}
%\begin{equation}
		\int_{-\infty}^{\infty}  \frac{\cos(at)-\cos(bt)}{t}h(t)\,dt > \\  \rho (b-a)+\rho \int_{-\infty}^{\infty}\frac{\sin(at)-\sin(bt)}{t}h(t)\,dt,
%\end{equation}
\end{multline}
and
\begin{multline}
%\begin{equation}
\int_{-\infty}^{\infty}\frac{\cos(ct)-\cos(dt)}{t}h(t)\,dt  < \\- \kappa\rho (d-c) - \kappa\rho \int_{-\infty}^{\infty}\frac{\sin(ct)-\sin(dt)}{t}h(t)\,dt.
%\end{equation}
\end{multline}
Summing the two inequalities, multiplied by $\lambda$ and $-\mu$ respectively, gives
%\begin{multline}
\begin{equation}
\label{eq:ct_ineq1}
\int_{-\infty}^{\infty} \psi(t)h(t)dt 
> \\  \rho [\lambda (b-a) +\kappa\mu(d-c)]
+\rho \int_{-\infty}^{\infty}\phi_1h(t)\,dt.
\end{equation}
%\end{multline}
%The inequality (\ref{main_inequality}) stands independent of conditions on $h(t)$. Megretski chooses $\lambda = r$ and $\mu=1/r$ to ensure $\psi(t)/\phi(t) \rightarrow 0$ as $t\rightarrow 0$.

\begin{enumerate}
\item[(i)] If $M\in\mathcal{M}^c$ then $\|h\|_1< 1$ and $h(t)\geq 0$ for all $t$. So
%\begin{multline}
\begin{multline}
\rho [\lambda (b-a) +\kappa\mu (d-c)] >\\  \int_{-\infty}^{\infty}\rho[\lambda(b-a) +\kappa\mu (d-c)] h(t)\,dt.
\end{multline}
%\end{multline}
and hence we can write (\ref{eq:ct_ineq1}) as 
\begin{equation}\label{eq:1a}
\int_{-\infty}^{\infty}  (\psi(t)-\rho \phi(t)) h(t)\,dt  > 0.
\end{equation}
But, since $\phi$ is an even function and $\psi$ is an odd function,
\begin{equation}
\psi(t)-\rho^c \phi(t) \leq 0\mbox{ for all }t.
\end{equation} 
Further, since $\phi$ is non-negative, 
\begin{equation}
\psi(t)-\rho \phi(t) \leq 0\mbox{ for all }t\mbox{ when }\rho\geq \rho^c.
\end{equation} 
Hence $\rho<\rho^c$.
\item[(ii)] If $M\in\mathcal{M}^c_{odd}$ then we can only say  $\|h\|_1< 1$. Nevertheless,
\begin{multline}
%\begin{equation}
\rho [\lambda (b-a) + \kappa\mu (d-c)] > \\ \int_{-\infty}^{\infty}\rho  [\lambda (b-a) + \kappa\mu (d-c)]  |h(t)|\,dt.
%\end{equation}
\end{multline}
and hence (\ref{eq:ct_ineq1}) leads to 
\begin{equation}\label{eq:6}
\int_{-\infty}^{\infty}  (|\psi(t)|-\rho \tilde{\phi}(t)) |h(t)|\,dt  > 0.
\end{equation}
But, since $\tilde{\phi}$ is also an even function and (as before) $\psi$ is an odd function,
\begin{equation}
|\psi(t)|-\rho^c_{odd} \tilde{\phi}(t) \leq 0\mbox{ for all }t.
\end{equation} 
Further, since $\tilde{\phi}$ is non-negative, 
\begin{equation}
|\psi(t)|-\rho \tilde{\phi}(t) \leq 0\mbox{ for all }t\mbox{ when }\rho\geq \rho^c_{odd}.
\end{equation} 
Hence $\rho<\rho^c_{odd}$.
\end{enumerate}

\subsection{Proof of Lemma~\ref{lemma:psi_dt}}
 The result is immediate following a similar argument to the proof of Lemma~\ref{lemma:psi}. In particular, as $\phi_d$ and $\tilde{\phi}_d$ are evaluated for discrete values of $n\geq 1$, their limiting behaviour as $n\rightarrow 0$ need not be considered.

\subsection{Proof of Theorem~\ref{thm:limit_dt}}

Suppose (\ref{fre1_dt}) holds for some multiplier $M(e^{j\omega})=1-H(e^{j\omega})$.
Then
\begin{equation}
\mbox{Im}(M(e^{j\omega}))  = \sum_{n=-\infty}^{\infty}\sin(\omega n) h_n
\end{equation}
and
\begin{equation}
\mbox{Re}(M(e^{j\omega}))  =1 -  \sum_{n=-\infty}^{\infty}\cos(\omega n) h_n,
\end{equation}
 where $h$ is the impulse response of $H$.
Hence integrating (\ref{fre1_dt})  over the interval $[a,b]$ gives
\begin{multline}
%\begin{equation}
\label{eq:dt_ineq1}
		\sum_{n=-\infty}^{\infty}  \frac{\cos(an)-\cos(bn)}{n}h_n > \\ \rho (b-a)+\rho \sum_{n=-\infty}^{\infty}\frac{\sin(an)-\sin(bn)}{n}h_n.
%\end{equation}
\end{multline}

\begin{enumerate}
\item[(i)] If $M\in\mathcal{M}^d$ then $h_0=0$, $\|h\|_1< 1$ and $h_n\geq 0$ for all $n$. So
\begin{equation}
\rho (b-a)  >  \sum_{n=-\infty}^{\infty}\rho(b-a)h_n,
\end{equation}
and hence we can write (\ref{eq:dt_ineq1}) as 
\begin{equation}\label{eq:1}
\sum_{n=-\infty}^{\infty}  (\psi_d(n)-\rho \phi_d(n)) h_n  > 0.
\end{equation}
But, since $\phi_d$ is an even function and $\psi_d$ is an odd function,
\begin{equation}
\psi_d(n)-\rho^d \phi_d(n) \leq 0\mbox{ for all }n\geq 1.
\end{equation} 
Further, since $\phi_d$ is non-negative, 
\begin{equation}
\psi_d(n)-\rho \phi_d(n) \leq 0\mbox{ for all }n\geq 1\mbox{ when }\rho\geq \rho^d.
\end{equation} 
Hence $\rho<\rho^d$.
\item[(ii)] If $M\in\mathcal{M}^d_{odd}$ then we can only say $h_0=0$ and  $\|h\|_1< 1$. Nevertheless,
\begin{equation}
\rho (b-a) > \sum_{n=-\infty}^{\infty}\rho (b-a)  |h_n|,
\end{equation}
and hence (\ref{eq:dt_ineq1}) leads to 
\begin{equation}\label{eq:6dt}
\sum_{n=-\infty}^{\infty}  (|\psi_d(n)|-\rho \tilde{\phi}_d(n)) |h_n|  > 0.
\end{equation}
But, since $\tilde{\phi}_d$ is also an even function and (as before) $\psi_d$ is an odd function,
\begin{equation}
|\psi_d(n)|-\rho^d_{odd} \tilde{\phi}_d(n) \leq 0\mbox{ for all }n\geq 1.
\end{equation} 
Further, since $\tilde{\phi}_d$ is non-negative, 
\begin{equation}
|\psi_d(n)|-\rho \tilde{\phi}_d(n) \leq 0\mbox{ for all }n\geq 1\mbox{ when }\rho\geq \rho^d_{odd}.
\end{equation} 
Hence $\rho<\rho^d_{odd}$.
\end{enumerate}

\subsection{Proof of Theorem~\ref{thm:int_dt}}

Substituting
\begin{equation}
\mbox{Im}(M(e^{j\omega}))  = \sum_{n=-\infty}^{\infty}\sin(\omega n) h_n
\end{equation}
and
\begin{equation}
\mbox{Re}(M(e^{j\omega}))  =1 -  \sum_{n=-\infty}^{\infty}\cos(\omega n) h_n,
\end{equation}
into (\ref{eq:int_dt})
leads to (\ref{eq:dt_ineq1}). The proof is then identical to that of Theorem~\ref{thm:limit_dt}.

\subsection{Proof of Proposition~\ref{thm:sparse}}$\mbox{ }$
\begin{enumerate}
\item[(i)]
Let $M_n(z)=1-z^{-n}$ with $n\in\mathcal{N}([a,b])$. Then
\begin{align}
	\mbox{Im}(M_n(e^{j\omega})) & = \sin(\omega n),\\
	\mbox{Re}(M_n(e^{j\omega})) & = 1 - \cos(\omega n).
\end{align}
Integrating over the interval yields 
\begin{equation}
\int_a^b\mbox{Im}(M(e^{j\omega}))\,d\omega = \rho^c \int_a^b \mbox{Re}(M(e^{j\omega}))\,d\omega.
\end{equation}
 Furthermore, if 
\begin{equation}
	M(z) = 1 - \sum_{n\in\mathcal{N}([a,b])} \lambda_n z^{-n},
\end{equation}
with
\begin{equation}
\lambda_n \geq 0 \mbox{ and }\sum_{n\in\mathcal{N}([a,b])} \lambda_n = 1,
\end{equation}
then we may write
\begin{equation}
	M(z) = \sum_{n\in\mathcal{N}([a,b])} M_n(z).
\end{equation}
The proof follows straightforwardly. 
\item[(ii)] Similar.
\end{enumerate}

\subsection{Proof of Lemma \ref{rho_bound_finite}}
Let
\begin{equation}
	\begin{aligned}
		\epsilon &= \frac{| \psi_d(1) |}{\phi_d(1)}= \frac{|\cos{a}-\cos{b}|}{b-a- (\sin{b}-\sin{a})} \\ 
	&= -\frac{\cos{b}-\cos{a}}{b-a- (\sin{b}-\sin{a})} = -\frac{\psi_d(1)}{\phi_d(1)},
\end{aligned}
\end{equation}
where we have used that $(x-\sin x)$ is a monotonically increasing function;
and
\begin{equation}
\begin{aligned}
	\nu & = \frac{2}{(b-a)}  \frac{1+\epsilon}{\epsilon} = \frac{2-2\psi_d(1)/\phi_d(1)}{- (b-a)\psi_d(1)/\phi_d(1)}  \\ 
	& = \frac{2(b-a)-2\sin b + 2\sin a -2 \cos b + 2 \cos a}{(a-b)(\cos b - \cos a)}.
\end{aligned}
\end{equation}
For $n> \nu$, we know $(b-a)n-2>0$. In addition, 
\begin{align}
	\epsilon (b-a)n > 2+2 \epsilon,
\end{align}
so 
\begin{align}
	\frac{| \psi_d(n) |}{\phi_d(n)} = \frac{| \cos{(bn)}-\cos{(an)} |}{(b-a)n - [\sin{(bn)}-\sin{(an)}]} < \frac{2}{(b-a)n - 2} < \epsilon.
\end{align}
As a result,
\begin{equation}
\frac{|\psi_d(n)|}{\phi_d(n)}<\frac{|\psi_d(1)|}{\phi_d(1)} \qquad \forall n>\nu.
\end{equation}

Finally, it is easy to check that 
\begin{equation}
\frac{|\psi_d(1)|}{\tilde{\phi}_d(1)}=
\frac{|\psi_d(1)|}{\phi_d(1)}
\end{equation}
and hence the same relation holds for $|\psi_d(n)|/\tilde{\phi}_d(n)$.

\bibliographystyle{IEEEtran}
\bibliography{IEEEabrv,Bibliography}

\begin{IEEEbiography}[{\includegraphics[width=1in,height=1.25in,clip,keepaspectratio]{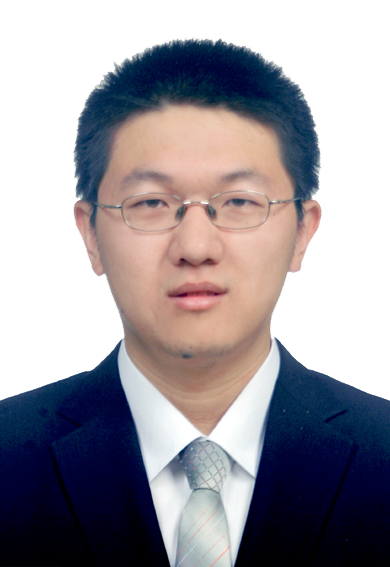}}]{Shuai Wang}
is a PhD candidate at the School of Electrical and Electronic Engineering, University of Manchester, UK. He received a Bachelor degree in Measurement and Control, Technology \& Instrument, and a Master degree in Instrument Science \& Technology from Harbin Institute of Technology, Harbin, China in 2011 and 2013 respectively. His research interests include absolute stability and multiplier theory. He is a member of IEEE, IET and SIAM.
\end{IEEEbiography}

\begin{IEEEbiography}[{\includegraphics[width=1in,height=1.25in,clip,keepaspectratio]{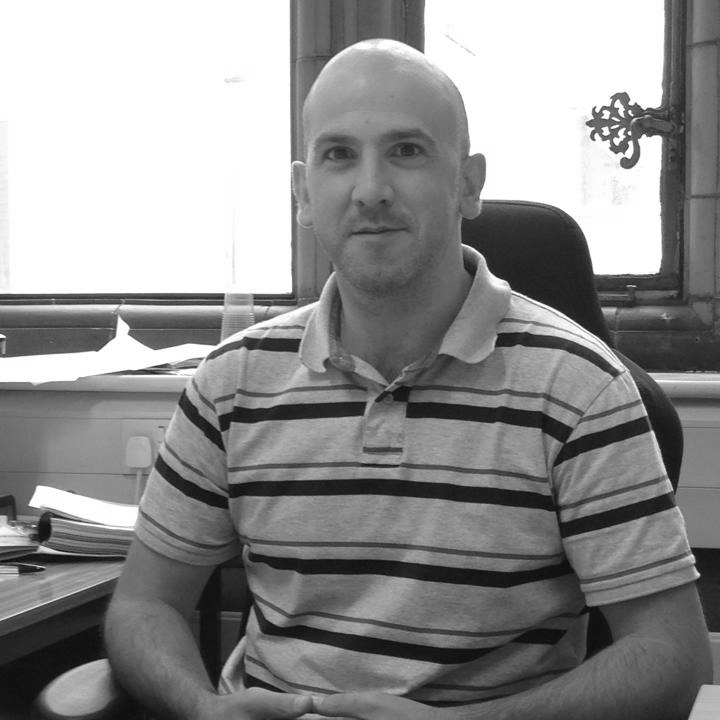}}]{Joaquin Carrasco}
is a Lecturer at the Control Systems Centre, School of Electrical and Electronic Engineering, University of Manchester, UK. He was born in Abarán, Spain, in 1978. He received the B.Sc. degree in physics and the Ph.D. degree in control engineering from the University of Murcia, Murcia, Spain, in 2004 and 2009, respectively. From 2009 to 2010, he was with the Institute of Measurement and Automatic Control, Leibniz Universität Hannover, Hannover, Germany. From 2010 to 2011, he was a research associate at the Control Systems Centre, School of Electrical and Electronic Engineering, University of Manchester, UK. He has been a Visiting Researcher at the University of Groningen, Groningen, The Netherlands, and the University of Massachusetts, Amherst. His current research interests include absolute stability, multiplier theory, and robotics applications. He is a member of the IFAC technical committee Telematics: Control via Communication Networks.
\end{IEEEbiography}

\begin{IEEEbiography}[{\includegraphics[width=1in,height=1.25in,clip,keepaspectratio]{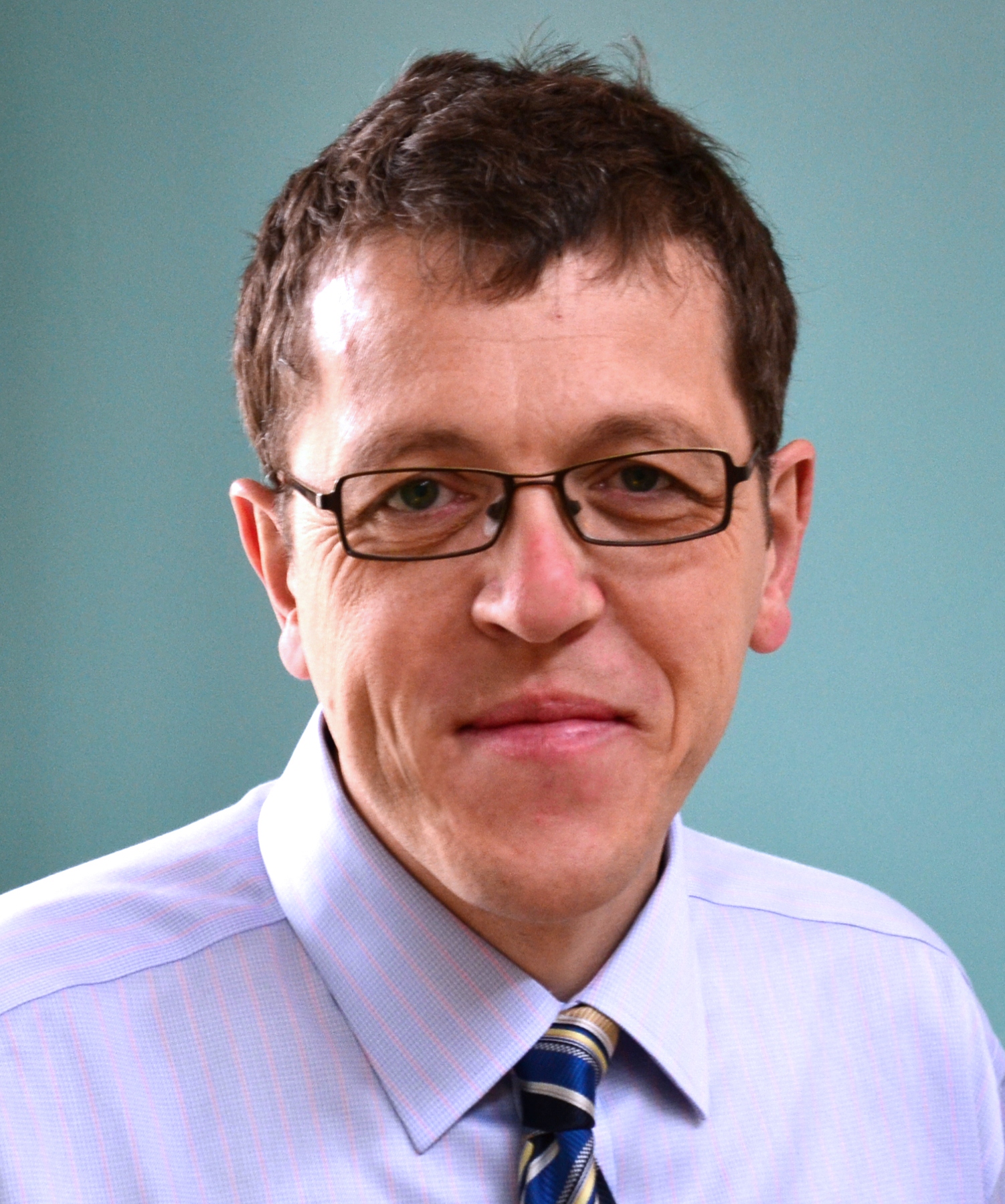}}]{William P. Heath}
is Chair of Feedback and Control at the School of Electrical and Electronic Engineering, University of Manchester, UK. He received a B.A. in Mathematics from Cambridge University in 1987, and both an M.Sc. and a Ph.D. from UMIST in 1989 and 1992 respectively. He was with Lucas Automotive from 1995 to 1998 and was a Research Academic at the University of Newcastle, Australia from 1998 to 2004. His interests include absolute stability, multiplier theory, constrained control and system identification.
\end{IEEEbiography}

\end{document}